\documentclass[vecphys]{svmult}

\usepackage{makeidx}         % allows index generation
\usepackage{graphicx}        % standard LaTeX graphics tool
\usepackage{multicol}        % used for the two-column index
\usepackage[bottom]{footmisc}% places footnotes at page bottom
\usepackage{amsmath}
\usepackage{amsfonts}
\usepackage{cite}
\DeclareMathOperator{\Tr}{Tr} \makeatletter
\DeclareRobustCommand\openone{\leavevmode\hbox{\small1\normalsize\kern-.33em1}}%

\makeindex             % used for the subject index
\begin{document}

\title*{Covariant Mappings for the Description of Measurement, Dissipation and
  Decoherence in Quantum Mechanics}
\titlerunning{Covariant Mappings in Quantum Mechanics}
\author{Bassano Vacchini\inst{1}}
% Use \authorrunning{Short Title} for an abbreviated version of
% your contribution title if the original one is too long
\institute{Dipartimento di Fisica
dell'Universit\`a di Milano and INFN,
Sezione di Milano
\\
Via Celoria 16, I--20133, Milan, Italy
\texttt{bassano.vacchini@mi.infn.it}}
%
% Use the package "url.sty" to avoid
% problems with special characters
% used in your e-mail or web address
%
\maketitle

\begin{abstract}
   The general formalism of quantum mechanics for the description of
   statistical experiments is briefly reviewed, introducing in
   particular position and momentum observables as POVM characterized
   by their covariance properties with respect to the isochronous
   Galilei group.  Mappings describing state transformations both as a
   consequence of measurement and of dynamical evolution for a closed
   or open system are considered with respect to the general constraints
   they have to obey and their covariance properties with respect to
   symmetry groups. In particular different master equations are
   analyzed in view of the related symmetry group, recalling the
   general structure of mappings covariant under the same
   group. This is done for damped harmonic oscillator, two-level
   system and quantum Brownian motion. Special attention is devoted to
   the general structure of translation-covariant master equations.
   Within this framework a recently obtained quantum counterpart of
   the classical linear Boltzmann equation is considered, as well as a
   general theoretical framework for the description of different
   decoherence experiments, pointing to a connection between different
   possible behaviours in the description of decoherence and the
   characteristic functions of classical L\'evy processes.
\end{abstract}

\section{Introduction}
\label{sec:introduction}

Since its very beginning quantum mechanics has urged physicists and
other scientists getting interested or involved with it to radically
change their classical picture of reality, as well as their way to
describe and understand experiments. Despite the elapsed time, the
matured knowledge about quantum mechanics and the growing number of
applications, the process of deeper understanding of quantum mechanics
and of its truly basic features is still on its way. In these notes we
will stress the standpoint that quantum mechanics actually is a
probability theory, enlarging and modifying the horizons of the
classical one and allowing to quantitative describe experiments at
microscopic level.  
The probabilistic standpoint aiming at an understanding of the
statistical structure of quantum theory has in fact turned out to be
of great significance in recent achievements in the description of
quantum mechanical systems, leading to the introduction of new
relevant concepts and tools often in analogy with classical probability theory.
In particular we will build on the growing impact
that the modern formulation of quantum mechanics in terms of
statistical operators, POVM and instruments is having in recent
experiments and theoretical studies, as well as the relevance of the
characterization of mappings giving the dynamical evolution of both
closed and open systems.  This applies as well to one step
transformations describing  the overall effect of a measurement, as
to measurements resolved in their time duration, or more generally
irreversible evolutions taking place as a consequence of the
interaction of the system of interest with some external system,
typically even though not necessarily having a very high number of
degrees of freedom. The presentation will pay particular attention to
the structural features of such mappings and especially to their
covariance with respect to the representation of a symmetry group,
whenever this applies.

The contribution grew out of the Lectures given at the Summer School in
Theoretical Physics in Durban and is organized as follows.
Sec.\ref{sec:quantum-mechanics-as} supports the point of view
according to which quantum mechanics actually is a probability theory
and its general formulation is presented in this spirit, starting from
the description of statistical experiments. States are introduced as
preparation procedures mathematically represented by statistical
operators, observables as registration procedures are described in
terms of POVM, the statistics of experimental outcomes is given in
terms of the trace formula. Finally general state transformations as a
consequence of measurement are associated to instruments, which
provide the transformed state as well as the statistics of the
measurement. Examples are provided focussing on position observables,
as well as joint position and momentum observables, understood as
specified according to their covariance properties with respect to the
isochronous Galilei group. Sec.\ref{sec:open-syst-covar} explores the
relevance of the concept of mapping acting on a space of operators in
a given Hilbert space for situations ranging from free evolutions to
open system dynamics. Dynamical mappings are characterized in view of
possible constraints depending on the physical situation of interest
and leading to important informations on their possible structure. In
particular we consider the notion of covariance with respect to
the representation of a symmetry group, complete positivity and
semigroup composition law, corresponding to a Markov approximation.
This leads to the Lindblad characterization of generators of
quantum-dynamical semigroups, possibly also including covariance
requirements. Particular examples of master equation are given
stressing their covariance properties with respect to the proper
symmetry, also providing the general characterization of
master equations covariant under the same group. This is done for the
damped harmonic oscillator and shift-covariance, a two-level system
and rotation-covariance, quantum Brownian motion and
translation-covariance. Finally it is shown how the general expression
of a translation-covariant generator, building on a quantum
non-commutative version of the L\'evy-Khintchine formula,
actually encompasses a quantum version of the classical linear
Boltzmann equation for the description of the motion of a quantum test
particle in a gas, as well as a unified theoretical framework for the
explanation of different decoherence experiments.

\section{Quantum mechanics as quantum probability}
\label{sec:quantum-mechanics-as}

In the present Section we will briefly introduce the basic tools
necessary in order to describe in the most general way a quantum
mechanical system and the possible measurements that can be performed
on it. The basic idea that we would like to convey or at least draw to
the reader's attention is that quantum mechanics is indeed naturally
to be seen as a probability theory, significantly different from the
classical one, rather than an extension of classical mechanics.
Experiments at the microscopic level are of statistical nature in an
essential way and their quantitative description asks for a
probabilistic model which is the quantum one, emerged in the 20's and
first thoroughly analyzed by von Neumann\cite{Neumann1932a}, actually
before the foundations of classical probability theory were laid down
by Kolmogorov in the 30's\cite{Kolmogorov1933}. The fact that quantum
mechanics is a probability theory different from the classical one,
containing the latter as a special case, brings with itself that
quantum experiments and their statistical description exhibit new
features, which sometimes appear unnatural or paradoxical when somehow
forced to fit in a classical probabilistic picture of reality, which
is closer to our intuition. Our presentation is more akin to the
introduction to quantum mechanics one finds in textbooks on quantum
information and communication theory rather than standard quantum
mechanics textbooks, even though at variance with the former we will
mainly draw examples from systems described in an infinite-dimensional
Hilbert space. The standpoint according to which quantum mechanics
actually is a probability theory is by now well understood, and even
though it is still not in the spirit of typical textbook
presentations, it has been developed and thoroughly investigated in
various books and monographes (see
e.g.\cite{LudwigFoundations1-LudwigFoundations2,Holevo1982,Kraus1983a,Streater2000a,Holevo2001a,Alicki2001a,Breuer2007,Strocchi2005}),
to which we refer the reader for more rigorous and detailed
presentations. A more concise account of similar ideas has also been
given in\cite{Vacchini2007a}.

\subsection{Quantum statistical description}
\label{sec:quant-stat-descr}

The basic setting of classical probability theory as clarified by
Kolmogorov is described within the mathematical framework of measure
theory. A classical probability model is fixed by specifying a measure
space, which is the space of elementary events, a $\sigma$-algebra on
this measure space characterizing the meaningful events to which we
want to ascribe probabilities, and a probability measure on it. The
observable quantities are then given by real measurable functions on
this space, i.e. random variables. Take for example the case of the
classical description of the dynamics of a point particle in
3-dimensional space. Then the measure space is given by the usual
phase-space $\mathbb{R}^3 \times \mathbb{R}^3$ endowed with the Borel
$\sigma$-algebra, where the points of phase-space can be identified
with position and momentum of the particle.  The probability measure
can be expressed by means of a probability density $f (\mathbf{x},\mathbf{p})$,
i.e. a positive and normalized element of $L^1 (\mathbb{R}^3 \times
\mathbb{R}^3)$ and observables are described as random variables given
by real functions $X (\mathbf{x},\mathbf{p})$ in $L^{\infty}
(\mathbb{R}^3 \times \mathbb{R}^3)$, so that exploiting the canonical
duality relation between $L^1$ and $L^{\infty}$ mean
values are given by
\begin{displaymath}
   \langle X\rangle_f = \int_{\mathbb{R}^3 \times \mathbb{R}^3} d^3 \! \mathbf{x}
    d^3 \! \mathbf{p} \,
X(\mathbf{x},\mathbf{p}) f(\mathbf{x},\mathbf{p}).
\end{displaymath}
The very same probability density $f(\mathbf{x},\mathbf{p})$ allows to
calculate the expectation value of any random variable, i.e. of any
observable. In particular any observable taking values in $\mathbb{R}$ defines a probability measure
on this space according to the formula
\begin{displaymath}
   \mu^{X}(M) =\int_{X^{-1}(M)} d^3 \! \mathbf{x}
    d^3 \! \mathbf{p} \,
f(\mathbf{x},\mathbf{p})
\end{displaymath}
where $M$ is a Borel set in the outcome space
$\mathbb{R}$ of the observable and again the same probability density
$f(\mathbf{x},\mathbf{p})$ appears. As a special case for the position
observable $X(\mathbf{x},\mathbf{p})=\mathbf{x}$ the measure
$\mu^{\mathbf{x}}$ can be expressed by means of the probability
density $f^{\mathbf{x}}(\mathbf{x})$ obtained by taking
the marginal of $f(\mathbf{x},\mathbf{p})$ with respect to momentum
and similarly for the momentum observable
$X(\mathbf{x},\mathbf{p})=\mathbf{p}$. In particular one can notice
that all observables commute, the points of phase-space can be taken
as meaningful elementary events and any probability measure can be
uniquely decomposed as a convex mixture of the extreme points of the
convex set of probability measures, given by the measures with support
concentrated at these elementary events given by single points in
phase-space. This probabilistic description is however not mandatory
in the classical case, where a deterministic description applies, and
it only becomes a very convenient or the only feasible tool for
systems with a very high number of degrees of freedom. The situation
is quite different in the quantum case.

\subsection{Statistics of an experiment}
\label{sec:stat-an-exper}

In the quantum case experiments are by necessity of statistical
nature. The most simple setup can be typically described as a suitably
devised macroscopic apparatus, possibly made up of lots of smaller
components, preparing the microscopical system we would like to study,
which in turn triggers another macroscopic device designed to measure
the value of a definite quantity. The reproducible quantity to be
compared with the theory is the relative frequency according to which
the preparation apparatus triggers the registration apparatus in a
high enough number of repetitions of the experiment under identical
circumstances. A most simple sketch of such a setup can be given by
the so-called Ludwig's Kisten\cite{LudwigFoundations1-LudwigFoundations2,Alber2001}
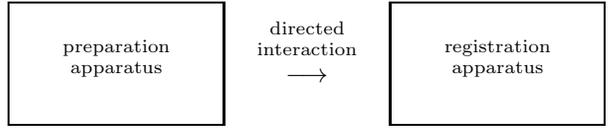
\begin{displaymath}
               \fbox{
            \hphantom{sp}
            \vrule height 20 pt depth 20 pt width 0 pt
    \mbox{$\mathrm{preparation}
    \atop\mathrm{apparatus}
    $}
            \vrule height 20 pt depth 20 pt width 0 pt
            \hphantom{sp}
            }
            \quad
{{\mathrm{directed}\atop\mathrm{interaction}}\atop \longrightarrow}
%    \xrightarrow[\mathrm{interaction}]{\mathrm{directed}}
            \quad
            \fbox{
            \hphantom{sp}
            \vrule height 20 pt depth 20 pt width 0 pt
    \mbox{$
    \textrm{registration}
    \atop
    \textrm{apparatus}
    $}
            \vrule height 20 pt depth 20 pt width 0 pt
            \hphantom{sp}
            }     \ .
\end{displaymath}
More complicated setups can be traced back to this one by suitably
putting together different apparata in order to build a new
preparation apparatus, and similarly for the registration part. In
order to describe such experiments one has to introduce a suitable
probability theory, which can actually account for the various
experimental evidences to be gained at microscopic level. This is
accomplished by introducing mathematical objects describing
the preparation and the registration, as well as a statistical formula
to extract from these two objects the probability densities to be
compared with the experimental outcomes.

\subsubsection{States as preparation procedures}
\label{sec:stat-as-prep}

In the quantum case a preparation procedure is generally described by
a statistical operator. Given the Hilbert space in which the system
one considers has to be described, e.g. $L^2 (\mathbb{R}^3)$ for the
centre of mass degrees of freedom of a particle in 3-dimensional
space, statistical operators are trace class operators positive and
with trace equal to one
\begin{displaymath}
    \rho\in\mathcal{K}(\mathcal{H})=\{\rho\in\mathcal{T}(\mathcal{H}) |
\rho=\rho^{\dagger}, \rho\geq 0, \Tr\rho=1\}.
\end{displaymath}
The set $\mathcal{K}(\mathcal{H})$ of statistical operators is a
convex subset of the space $\mathcal{T}(\mathcal{H})$ of trace class
operators on the Hilbert space $\mathcal{H}$, so that any convex
mixture of statistical operators is again a statistical operator
\begin{displaymath}
   \rho_1,\rho_2\in\mathcal{K}(\mathcal{H}) \Rightarrow   w=\mu{\rho}_{1}+
  (1-\mu){\rho}_{2}\in\mathcal{K}(\mathcal{H}) \quad 0\leq \mu \leq 1 .
\end{displaymath}
In particular the extreme points of such a  set are given by one
dimensional projections, that is to say pure states, which
cannot be expressed as a proper mixture. We stress the fact that
statistical operators are actually to be associated to the considered
statistical preparation procedure, rather than to the system
itself. More precisely they describe a whole equivalence class of
preparation procedures which all prepare the system in the same state,
even though by means of quite different macroscopic apparata. This
correspondence between statistical operators and equivalence classes
of preparation procedures is reflected in the fact that a statistical
operator generally admits infinite different decompositions, as
mixtures of pure states or other statistical operators. Relying on the
spectral theorem a statistical operator can always be written in the form
\begin{displaymath}
   {\rho}=\sum_j {\lambda_j}|\psi_j\rangle\langle\psi_j|, \quad
\lambda_j\geq 0, \quad \sum_j {\lambda_j}=1
\end{displaymath}
with $\{ |\psi_j\rangle\}$ an orthonormal set. However
infinite others not necessarily orthogonal demixtures generally
exist. Think for example of the most simple case of a statistical
operator describing the spin state of a fully unpolarized beam of spin
$1/2$ particles: $\rho=\frac{1}{2}\openone$. Then any orthogonal basis
in $\mathbb{C}^2$ (e.g. the eigenvectors of the spin operator along a
given, arbitrary direction) does provide an orthogonal decomposition
of the considered statistical operator. Such decompositions do
correspond to different possible macroscopic procedures leading to
such a preparation. In the $N$ runs of the statistical experiment the
beam is prepared $\frac{N}{2}$ times with spin $+\frac{\hbar}{2}$
along a fixed direction, and $\frac{N}{2}$ times with spin
$-\frac{\hbar}{2}$. All such preparations, differing in the choice of direction, lead to the same state, but
they cannot be performed together since no apparatus can measure the
spin along two different directions: they are therefore
incompatible. The prepared state is however the same and the actual
preparation cannot be distinguished on the basis of any other subsequent
statistical experiment whatsoever performed on the obtained state. At
variance with the classical case therefore, states are given by
operators which generally admit infinite convex decompositions, and
represent equivalence class of preparation procedures.

\subsubsection{Observables as registration procedures}
\label{sec:observ-as-registr}

On the same footing one has to associate a mathematical object to a
macroscopic apparatus assembled in order to measure the value of a
certain quantity. Once again utterly different and generally
incompatible macroscopic procedures and apparata can possibly be used
to assign a value to the same physical quantity. The operator
describing an observable is therefore to be understood as the
mathematical representative of a whole equivalence class of
registration procedures. In full generality an observable in the sense
clarified above is given by a positive operator-valued measure
(POVM), the measure theoretic aspect appearing since one is in fact
interested in the probability that the quantity of interest does lie
within a certain interval. 

A POVM is a mapping defined on a suitable
measure space and taking values in the set of positive operators
within $\mathcal{B}(\mathcal{H})$, that is to say the Banach space of
bounded operators on $\mathcal{H}$. Taking for the sake of
concreteness an observable assuming values in $\mathbb{R}^3$, such as
the position of a particle in 3-dimensional space, a POVM is given by
a mapping $F$ defined on the Borel $\sigma$-algebra $\mathcal{B}
(\mathbb{R}^3)$.
\begin{align*}
   F:\mathcal{B} (\mathbb{R}^3) &\rightarrow \mathcal{B} (\mathcal{H})
\\
M &\rightarrow F (M)
\end{align*}
associating to each interval $M\in \mathcal{B} (\mathbb{R}^3)$ a
positive bounded operator in such a way that
\begin{flalign*}
& 0\leq F (M)\leq \openone
\\
& F (\emptyset)=0
\qquad  F (\mathbb{R}^3)=\openone
\\
& F (\cup_{i} M_i)=\sum_{i} F (M_i) \qquad M_i\cap M_j=\emptyset \quad
i\not = j
\end{flalign*}
where the first condition will turn out to be necessary for the
statistical interpretation, the second one expresses normalization,
always associating the null operator to the empty set and the identity
to the whole space, while the last condition amounts to
$\sigma$-additivity. For fixed set $M\in \mathcal{B} (\mathbb{R}^3)$
the operator $F (M)$, positive and between zero and one, is called
effect. Note that observables are here given by generally non
commuting operators. Moreover we have not requested $F (M)$ to be a
projection operator, that is to say a self-adjoint and idempotent
operator such that $F^2 (M)=F (M)$. If this further condition holds
for all $ M\in \mathcal{B} (\mathbb{R}^3)$ one has a very special case
of POVM, also called projection-valued measure (PVM), since it is a
measure taking values in the space of projections on the Hilbert space
$\mathcal{H}$. For such measures we shall use the symbol $E (M)$. In
the case of PVM there is a one to one correspondence between the PVM
and a uniquely defined self-adjoint operator, thus explaining the
standard definition of observable as self-adjoint operator. The
operator associated to the PVM turns out to be a very convenient tool
for the calculation of mean values and higher order moments, such as
variances. 
Let us call $E^{\mathbf{A}}$ the PVM for the description of
measurements on the quantity $\mathbf{A}$ taking values in $\mathbb{R}^k$. The first moment of the measure
\begin{displaymath}
   \mathbf{A}=\int_{\sigma (\mathbf{A})} \mathbf{x} dE^{\mathbf{A}}(\mathbf{x})
\end{displaymath}
actually identifies $k$ commuting self-adjoint operators, the integral being
calculated over the support of the measure or equivalently the
spectrum of $\mathbf{A}$, and higher moments of the measure can be
identified with powers of these operators, according to the functional calculus
\begin{displaymath}
      \mathbf{A}^{n}=\int_{\sigma (\mathbf{A})} \mathbf{x}^{n} dE^{\mathbf{A}}(\mathbf{x}).
\end{displaymath}
In particular a whole collection of commuting self-adjoint operators
can be obtained considering the integrals of a measurable function
$g$ from $\mathbb{R}^k$ to $\mathbb{R}$
\begin{displaymath}
      g (\mathbf{A})=\int_{\sigma (\mathbf{A})} g (\mathbf{x}) dE^{\mathbf{A}}(\mathbf{x}),
\end{displaymath}
corresponding to measurements of functions of the quantity $\mathbf{A}$.
These facts are no longer true for a generic POVM.

\subsubsection{Statistics of outcomes}
\label{sec:statistics-outcomes}

Having introduced statistical operators as general mathematical
representatives of a state, in the sense of characterization of 
preparation apparata, and POVM as mathematical representatives of
observable, associated to registration apparata, we now have to
combine states and observables in order to express the probabilities to
be compared with the outcomes of an experiment. This is done by
considering the duality relation between the spaces
of states and observables. As in the classical description the space of
observables $L^{\infty}$ was the dual of the space $L^{1}$ of states,
here the Banach space of bounded operators  is the dual of the space
$\mathcal{T}(\mathcal{H})$ of trace class operators to which
statistical operators do belong. The duality relation is given by the
trace evaluation
\begin{align*}
   \Tr: \mathcal{B} (\mathcal{H}) \times  \mathcal{T} (\mathcal{H})&\rightarrow \mathbb{C}
\\
(X,w) &\rightarrow \Tr X^{\dagger}w,
\end{align*}
where taking any basis in $\mathcal{H}$, e.g. $\{u_n \}$, the trace
can be evaluated as
\begin{displaymath}
   \Tr X^{\dagger}w=\sum_{n}\langle u_n |X^{\dagger}w |u_n\rangle
\end{displaymath}
the series being convergent for any bounded operator $X$ and trace
class operator $w$ and the result independent of the choice of
orthonormal basis. Given a system prepared in the state $\rho$, the
probability that a quantity described by the POVM $F$ takes value in
the set $M$ is given by the statistical formula
\begin{equation}
   \label{eq:1}\Tr\rho F (M).
   \end{equation}
The property of $\rho$ and $F$ ensure that $\Tr\rho F (M)$ is indeed a
positive number between zero and one, and in particular for every
couple $\rho$ and $F$ the mapping
\begin{align*}
   \Tr\rho F (\cdot):\mathcal{B} (\mathbb{R}^3) &\rightarrow [0,1]
\\
M &\rightarrow  \Tr\rho F (M)
\end{align*}
is a classical probability measure assigning to each set $M$ the
probability $\Tr\rho F (M) $ that the outcome of the experiment lies in
that set. For a given state to each observable one can therefore
associate a classical probability measure, however only commuting
observables are described by the same probability measure, to
different observables one generally has to associate distinct
probability measures. The formula~\eqref{eq:1} when considered for the
particular case of a pure state $|\psi\rangle$ and a PVM $E$ leads to
the usual expression
\begin{displaymath}
   \| E (M)\psi\|^2
\end{displaymath}
for the evaluation of the statistics of an experiment measuring $E$
once the system has been prepared in the state $|\psi\rangle$.
Considering within this framework the usual notion of position and
momentum observables one immediately realizes that the related
measures can be expressed by means of the two well distinct
probability densities $|\psi(\mathbf{x})|^2$ and
$|\tilde\psi(\mathbf{p})|^2$ respectively ($\tilde\psi$ denoting as
usual the Fourier transform). At variance with the classical case
there generally is no common probability density allowing to express
the probability measure of all observables. There is in fact no sample
space of elementary events. Note that for fixed $F$ the mapping
$\rho\rightarrow \Tr\rho F (\cdot)$ is an affine mapping from the
convex set $\mathcal{K}(\mathcal{H})$ of statistical operators into the
convex set of classical probability measures on $\mathcal{B}
(\mathbb{R}^3)$. For fixed observable this is all we need in order to
compare with the experimental outcomes. Since any such affine mapping
can be written in this form for a uniquely defined POVM one can come
to the conclusion that POVM indeed provide the most general
description of the statistics of experimental outcomes compatible with
the probabilistic interpretation of quantum mechanics. The statistical
formula~\eqref{eq:1} is the key point where the theory can be compared
with experiment and allows us to better understand the meaning of
equivalence class. Two preparation apparata are in the same
equivalence class if they produce the very same statistics of outcomes
for any observable, and similarly two registration apparata are in the
same equivalence class if they lead to the same statistics of outcomes
for any state.

\subsection{Example of POV for position and momentum}
\label{sec:example-pov-position}

We now want to consider a few examples of POVM and PVM concentrating
on position and momentum, showing in particular how symmetry
properties can be a very important guiding principle in the
determination of meaningful observables, once we leave the
correspondence principle focussed on quantum mechanics as a new
mechanics with respect to the classical one. As we shall see while for
the case of either position or momentum alone POVM are essentially
given by a suitable coarse-graining with respect to the usual PVM, if
one wants to give statistical predictions for the measurement of both
position and momentum together the corresponding observable is given
by necessity in terms of a POVM. For a more detailed and
mathematically accurate exposition we refer the reader
to\cite{Davies1976a,Busch1995a,Holevo2001a}.

\subsubsection{Covariant mapping}
\label{sec:covariant-mapping}

Let us first start by introducing the notion of mapping covariant
under a given symmetry group $G$. As we will show this notion is of
great interest in many situations, both for the construction of POVM and
general dynamical mappings. Consider a measure space $\mathcal{X}$
with a Borel $\sigma$-algebra of sets $\mathcal{B}
(\mathcal{X})$. Such a space is called a $G$-space if there exist an
action of $G$ on $\mathcal{X}$ defined as a mapping that sends group
elements $g\in G$ to transformation mappings $\mu_g$ on $\mathcal{X}$
in such a way as to preserve group composition and identity
\begin{displaymath}
\mu_g   \mu_h=\mu_{gh} \qquad \forall g,h\in G \qquad \mu_e=1_{\mathcal{X}}
\end{displaymath}
where $1_{\mathcal{X}}$ denotes the identity function on
$\mathcal{X}$. If furthermore $G$ acts transitively on $\mathcal{X}$,
in the sense that any two point of $\mathcal{X}$ can be mapped one
into the other with $\mu_{g'}$ for a suitable $g'\in G$, then
$\mathcal{X}$ is called a transitive $G$-space . Consider for example
$\mathcal{X}=\mathbb{R}^3$, then $\mathcal{X}$ is a transitive
$G$-space with respect to the group of translations. The elements of
the group are 3-dimensional vectors acting in the obvious way on the
Borel sets of $\mathbb{R}^3$, i.e. $\mu_{\mathbf{a}}=M+\mathbf{a}$ for
all
$ \mathbf{a}\in\mathbb{R}^3$ and for all $M\in \mathcal{B}
(\mathbb{R}^3) $. Consider as well a unitary representation $U (g)$ of
the same group $G$ on a Hilbert space $\mathcal{H}$
\begin{displaymath}
   |\psi_g\rangle=U (g)|\psi\rangle \qquad \psi\in\mathcal{H}\qquad
   g\in G,
\end{displaymath}
in terms of which one also has a representation of $G$ on a
space $\mathcal{A} (\mathcal{H})$ of operators acting on $\mathcal{H}$
\begin{displaymath}
   A_g=U^{\dagger} (g)A U (g) \qquad A\in\mathcal{A} (\mathcal{H}).
\end{displaymath}
A mapping $\mathcal{M}$ defined on $\mathcal{B}
(\mathcal{X})$ and taking values in $\mathcal{A} (\mathcal{H})$ is
said to be covariant with respect to the symmetry group $G$ provided
it commutes with the action of the group in the sense that
\begin{equation}
\label{cov}
   U^{\dagger} (g)\mathcal{M} (X) U (g)=\mathcal{M} (\mu_{g^{-1}} (X))
   \qquad \forall X\in\mathcal{B}
(\mathcal{X}) \quad \forall g\in G.
\end{equation}
A symmetry transformation on the domain of the mapping is mapped into
the symmetry transformation corresponding to the same group element on
the range of the mapping.

\subsubsection{Position observable}
\label{sec:position-observable}

As an example of observable in the sense outlined above we now want to
introduce the position observable. Rather than relying on the usual
correspondence principle with respect to classical mechanics, we want
to give an operational definition of the position observable, fixing
its behaviour with respect to the action of the relevant symmetry
group, which in this case is the isochronous Galilei group, containing
translations, rotations and boosts that is to say velocity
transformations. The group acts in the natural way on the Borel sets
of $\mathbb{R}^3$, and the covariance equations that we require for an
observable to be interpreted as position observable are the following
\begin{align}
\label{covx}
\nonumber
         U^{\dagger} (\mathbf{a})F^{\mathbf{x}} (M) U
         (\mathbf{a})&=F^{\mathbf{x}} (M-\mathbf{a}) && \forall \mathbf{a}\in\mathbb{R}^3
         \\
         U^{\dagger} (\mathsf{R})F^{\mathbf{x}} (M) U
         (\mathsf{R})&=F^{\mathbf{x}} (\mathsf{R}^{-1}M)&& \forall \mathsf{R}\in SO (3)
         \\
\nonumber
         U^{\dagger} (\mathbf{q})F^{\mathbf{x}} (M) U
         (\mathbf{q})&=F^{\mathbf{x}} (M)&& \forall \mathbf{q}\in\mathbb{R}^3.
\end{align}
The mapping $F^{\mathbf{x}}$ to be interpreted as a position
observable has to transform covariantly with respect to translations
and rotations as in~\eqref{cov}, and to be invariant under a velocity
transformation. These equations can also be seen as a requirement on
the possible macroscopic apparata possibly performing such a
measurement. The apparatus used to test whether the considered system
is localized in the translated region $M-\mathbf{a}$ should be in the
equivalence class to which the translated  apparatus used to test
localization in the region $M$ belongs, and similarly for
rotations. Localization measurements should instead be unaffected by
boost transformations. A solution of these covariance equations, that
is to say a POVM complying with~\eqref{covx}, is now a position
observable. If one looks for such a solution asking moreover that the
POVM be in particular a PVM, the solution is uniquely given by the
usual spectral decomposition of the position operator
\begin{equation}
   \label{ex}
   E^{\mathbf{x}} (M)=\chi_{M} (\hat{\mathbf{x}}) =\int_{M}d^3\!\mathbf{x} |\mathbf{x}\rangle\langle\mathbf{x}|
\end{equation}
where $\chi_{M}$ denotes the characteristic function of the set
$M$. The first moment of the spectral measure gives the usual triple
of commuting position operators
\begin{displaymath}
   \hat{\mathbf{x}} =\int_{\mathbb{R}^3}d^3\!\mathbf{x}\,\mathbf{x} |\mathbf{x}\rangle\langle\mathbf{x}|,
\end{displaymath}
whose powers coincide with the higher moments of $E^{\mathbf{x}}$
\begin{displaymath}
   \hat{\mathbf{x}}^n =\int_{\mathbb{R}^3}d^3\!\mathbf{x}\,\mathbf{x}^{n} |\mathbf{x}\rangle\langle\mathbf{x}|.
\end{displaymath}
In particular for a given state $\rho$ mean values and variances of
the classical probability distribution giving the position
distribution can be expressed by means of
the operator $\hat{\mathbf{x}}$
\begin{align*}
   \textsf{\textsf{Mean}} (E^{\mathbf{x}})&= \Tr\rho \hat{\mathbf{x}}=  \langle
   \hat{\mathbf{x}}\rangle_{\rho}
\\
\textsf{Var}   (E^{\mathbf{x}})&= \Tr\rho \hat{\mathbf{x}}^2 - (\Tr\rho
\hat{\mathbf{x}})^2 =\langle \hat{\mathbf{x}}^2\rangle_{\rho}
- \langle \hat{\mathbf{x}}\rangle_{\rho}^2.
\end{align*}
The couple $(U,E^{\mathbf{x}})$, where $U$ is the unitary
representation of the symmetry group, here the isochronous Galilei
group, and $E^{\mathbf{x}}$ a PVM covariant under the action of $U$ is
called a system of imprimitivity.
More generally a solution of~\eqref{covx} as a POVM is obtained as
follows. Let us introduce a rotationally invariant probability density
$h (\mathbf{x})$
\begin{displaymath}
     h (\mathbf{x})\geq0 \qquad \int d^3\!\mathbf{x}\,h
         (\mathbf{x})=1
         \qquad
         h (\mathsf{R}\mathbf{x})=h (\mathbf{x}),
\end{displaymath}
with 
% zero mean for the sake of simplicity
% \begin{displaymath}
%    \textsf{Mean} (h)=\int d^3\!\mathbf{x}\, \mathbf{x}h
%          (\mathbf{x})=0
% \end{displaymath}
% and therefore 
variance given by
\begin{displaymath}
   \textsf{Var} (h)=\int d^3\!\mathbf{x}\, \mathbf{x}^2 h
         (\mathbf{x}).
\end{displaymath}
One can then indeed check that the expression
\begin{equation}
   \label{fx}
   F^{\mathbf{x}} (M)= (\chi_{M}*h) (\hat{\mathbf{x}})
         =\int_{M} d^3\!\mathbf{y}\int_{\mathbb{R}^3} d^3\!\mathbf{x}\, h (\mathbf{x}-\mathbf{y}) |\mathbf{x}\rangle\langle\mathbf{x}|
\end{equation}
where $*$ denotes convolution, actually is a POVM complying
with~\eqref{covx}, and in fact provides the general solution
of~\eqref{covx}. The POVM~\eqref{fx} actually is a smeared version of
the usual sharp position observable, the probability density $h
(\mathbf{x})$ which fixes the POVM being understood as the actual,
finite resolution of the registration apparatus. For any state $\rho$
the first moment of the associated probability density can still be
expressed as the mean value of the usual position operator, since $\textsf{Mean} (h)=0$
\begin{align*}
   \textsf{Mean} (F^{\mathbf{x}})&=\int_{\mathbb{R}^3}
   d^3\!\mathbf{y}\int_{\mathbb{R}^3} d^3\!\mathbf{x}\, \mathbf{y} h
   (\mathbf{x}-\mathbf{y}) \Tr \rho |\mathbf{x}\rangle\langle\mathbf{x}|
\\
&=\langle \hat{\mathbf{x}}\rangle_{\rho}
\end{align*}
the second moment however differs
\begin{align*}
   \textsf{Var} (F^{\mathbf{x}})&=\int_{\mathbb{R}^3}
   d^3\!\mathbf{y}\int_{\mathbb{R}^3} d^3\!\mathbf{x}\, \mathbf{y}^2 h
   (\mathbf{x}-\mathbf{y}) \Tr \rho
   |\mathbf{x}\rangle\langle\mathbf{x}|
\\
&
 \hphantom{=\int_{\mathbb{R}^3}
    d^3\!\mathbf{y}\int_{\mathbb{R}^3} d^3\!\mathbf{x}\, \mathbf{y}^2 }
- \left(\int_{\mathbb{R}^3}
   d^3\!\mathbf{y}\int_{\mathbb{R}^3} d^3\!\mathbf{x}\, \mathbf{y} h
   (\mathbf{x}-\mathbf{y}) \Tr \rho |\mathbf{x}\rangle\langle\mathbf{x}|\right)^2
\\
&=\langle \hat{\mathbf{x}}^2\rangle_{\rho}- \langle \hat{\mathbf{x}}\rangle_{\rho}^2+\textsf{Var} (h),
\end{align*}
it is no more expressed only by the mean value of the operator which
can be used to evaluate the first moment and by its square. A further
contribution $\textsf{Var} (h)$ appears, which is state independent and
reflects the finite resolution of the equivalence class of apparata
used for the localization measurement. Note that the usual result is
recovered in the limit of a sharply peaked probability density $h
(\mathbf{x}) \rightarrow \delta^3(\mathbf{x})$. Taking e.g. a
distribution of the form
\begin{displaymath}
   h_{\sigma}(\mathbf{x})=\left (\frac{1}{2\pi
     \sigma^2}\right)^{\frac{3}{2}}e^{-\frac{1}{2\sigma^2}\mathbf{x}^2}
\xrightarrow{\sigma \rightarrow 0}\delta^3(\mathbf{x})
\end{displaymath}
one has that in the limit of an infinite accuracy in the localization
measurement of the apparatus exploited the POVM reduces to the
standard PVM
\begin{align*}
   F^{\mathbf{x}} (M)&=\int_{M} d^3\!\mathbf{y}\int_{\mathbb{R}^3}
   d^3\!\mathbf{x}\,
\left (\frac{1}{2\pi
     \sigma^2}\right)^{\frac{3}{2}}
e^{-\frac{1}{2\sigma^2}(\mathbf{x}-\mathbf{y})^2}|\mathbf{x}\rangle\langle\mathbf{x}|
\\
&\xrightarrow{\sigma \rightarrow 0} \int_{M}
d^3\!\mathbf{y}\int_{\mathbb{R}^3} d^3\!\mathbf{x}\,
\delta^3(\mathbf{x}-\mathbf{y})|\mathbf{x}\rangle\langle\mathbf{x}|
\\
&=\int_{M} d^3\!\mathbf{x}\,|\mathbf{x}\rangle\langle\mathbf{x}|.
\end{align*}
Analogous results can obviously be obtained for a momentum observable,
asking for the corresponding covariance properties.

\subsubsection{Position and momentum observable}
\label{sec:posit-moment-observ}

A more interesting situation appears when considering apparata
performing both a measurement of the spatial location of a particle as
well as of its momentum. As it is well known no observable can be
associated to such a measurement in the framework of standard textbook
quantum mechanics. Let us consider the covariance equations of such an
observable in the more general framework of POVM. A position and
momentum observable should be given by a POVM
$F^{\mathbf{x},\mathbf{p}}$ defined on $\mathcal{B}
(\mathbb{R}^3 \times \mathbb{R}^3)$ satisfying the following
covariance equations under the action of translations, rotations and
boosts respectively
\begin{align}
\label{covxp}
\nonumber
   U^{\dagger} (\mathbf{a})F^{\mathbf{x},\mathbf{p}} (M\times N) U
   (\mathbf{a})&=F^{\mathbf{x},\mathbf{p}} (M-\mathbf{a}\times N) &&\forall \mathbf{a}\in\mathbb{R}^3
\\
   U^{\dagger} (\mathsf{R})F^{\mathbf{x},\mathbf{p}} (M\times N) U
   (\mathsf{R})&=F^{\mathbf{x},\mathbf{p}} (\mathsf{R}^{-1}M\times \mathsf{R}^{-1} N) &&\forall \mathsf{R}\in SO (3)
\\
\nonumber
   U^{\dagger} (\mathbf{q})F^{\mathbf{x},\mathbf{p}} (M\times N) U
   (\mathbf{q})&=F^{\mathbf{x},\mathbf{p}} (M\times N-\mathbf{q}) &&\forall \mathbf{q}\in\mathbb{R}^3.
\end{align}
Such covariance equations, defining a position and momentum observable
by means of its operational meaning, do not admit any solution within
the set of PVM, while the general solution within the set of POVM is
given by
\begin{equation}
\label{fxp}
           F^{\mathbf{x},\mathbf{p}} (M\times N)
           =\frac{1}{(2\pi\hbar)^3}\int_{M}
           d^3\!\mathbf{x}\int_{N}d^3\!\mathbf{p}
\,
           W  (\mathbf{x},\mathbf{p}) S W^{\dagger}(\mathbf{x},\mathbf{p}) 
        \end{equation}
where $S$ is a trace class operator, positive, with trace equal to one
and invariant under rotations
\begin{displaymath}
   S\in\mathcal{T}(\mathcal{H})\quad S\geq 0 \quad \Tr S=1 \qquad
           U^{\dagger} (\mathsf{R}) S U (\mathsf{R})=S
\end{displaymath}
so that it is in fact a statistical operator, even though it does not
have the meaning of a state, while the unitaries
\begin{displaymath}
   W (\mathbf{x},\mathbf{p}) =e^{-\frac{i}{\hbar}
             (\mathbf{x}\cdot\hat{\mathbf{p}} -\hat{\mathbf{x}} \cdot
             \mathbf{p} )} 
\end{displaymath}
are the Weyl operators built in terms of the canonical position and
momentum operators. The covariance of~\eqref{fxp} under \eqref{covxp}
can be directly checked, together with its normalization, working
with the matrix elements of the operator expression. The couple
$(U,F^{\mathbf{x},\mathbf{p}})$, where $U$ is the unitary
representation of the symmetry group and $F^{\mathbf{x},\mathbf{p}}$ a
POVM covariant under its action is now called system of covariance.
The connection with position and momentum observable as well as the
reason why such a joint observable can be expressed only in the
formalism of POVM, where position observables alone are generally
given by smeared versions of the usual  position observable, and
similarly for momentum, can be understood  looking at the marginal
observables. Starting from~\eqref{fxp} one can in fact consider a
measure of position irrespective of the momentum of the particle,
thus coming to the marginal position observable
\begin{align*}
   F^{\mathbf{x}} (M) &= F^{\mathbf{x},\mathbf{p}} (M\times
   \mathbb{R}^3)
\\
&=\int_{M}d^3\!\mathbf{y}\int_{\mathbb{R}^3} d^3\!\mathbf{x}\,  |\mathbf{x}\rangle
\langle \mathbf{x}-\mathbf{y}|S| \mathbf{x}-\mathbf{y}\rangle \langle\mathbf{x}|
\\
&=\int_{M}d^3\!\mathbf{y}\int_{\mathbb{R}^3} d^3\!\mathbf{x}\, h_{S^{\mathbf{x}}}
         (\mathbf{x}-\mathbf{y}) |\mathbf{x}\rangle\langle\mathbf{x}|
\end{align*}
where the function
\begin{equation}
   \label{hx}
   h_{S^{\mathbf{x}}}
         (\mathbf{x})=\langle \mathbf{x}|S| \mathbf{x}\rangle 
\end{equation}
is a well defined probability density due to the fact that the
operator $S$ has all the properties of a statistical operator, so that
$\langle \mathbf{x}|S| \mathbf{x}\rangle$ would be the position
probability density of a system described by the state $S$. On similar
grounds the marginal momentum observable is given by
\begin{align*}
   F^{\mathbf{p}} (M) &= F^{\mathbf{x},\mathbf{p}} (\mathbb{R}^3\times
   N)
\\
&=\int_{N}d^3\!\mathbf{q}\int_{\mathbb{R}^3} d^3\!\mathbf{p}\,  |\mathbf{p}\rangle
\langle \mathbf{p}-\mathbf{q}|S| \mathbf{p}-\mathbf{q}\rangle \langle\mathbf{p}|
\\
&=\int_{N}d^3\!\mathbf{q}\int_{\mathbb{R}^3} d^3\!\mathbf{p}\, h_{S^{\mathbf{p}}}
         (\mathbf{p}-\mathbf{q}) |\mathbf{p}\rangle\langle\mathbf{p}|
\end{align*}
where again the function
\begin{equation}
   \label{hp}
   h_{S^{\mathbf{p}}}
         (\mathbf{p})=\langle \mathbf{p}|S| \mathbf{p}\rangle 
\end{equation}
is a well defined probability density, which corresponds to the
momentum probability density of a system described by the statistical
operator $S$.  As it appears the marginal observables are given by two
POVM characterized by a smearing of the standard position and momentum
observables by means of the probability densities $h_{S^{\mathbf{x}}}
(\mathbf{x})$ and $h_{S^{\mathbf{p}}} (\mathbf{p})$ respectively. It
is exactly this finite resolution in the measurement of both position
and momentum, with two probability densities satisfying
\begin{displaymath}
   \textsf{Var}_{i} (h_{S^{\mathbf{x}}})\textsf{Var}_{i}
   (h_{S^{\mathbf{p}}})\geq \frac{\hbar^2}{4}
\qquad i=x,y,z
\end{displaymath}
as follows from~\eqref{hx} and~\eqref{hp}, that allows for a joint
measurement for position and momentum in quantum mechanics, without
violating Heisenberg's uncertainty relations. In order to consider a
definite example we take $S$ to be a pure state corresponding to a
Gaussian of width $\sigma$
\begin{displaymath}
   \langle \mathbf{x}|\psi\rangle = 
\left (\frac{1}{2\pi
     \sigma^2}\right)^{\frac{3}{4}}
e^{-\frac{1}{4\sigma^2}\mathbf{x}^2},
\end{displaymath}
on which the Weyl operators act as a translation in both position and
momentum, leading to
\begin{displaymath}
   \langle \mathbf{x}|W (\mathbf{x}_0,\mathbf{p}_0)|\psi\rangle = \left (\frac{1}{2\pi
     \sigma^2}\right)^{\frac{3}{4}}e^{-\frac{1}{4\sigma^2}(\mathbf{x}-\mathbf{x}_0)^2+\frac{i}{\hbar}\mathbf{p}_0\cdot (\mathbf{x}-\mathbf{x}_0)}=\langle \mathbf{x}|\psi_{\mathbf{x}_0,\mathbf{p}_0}\rangle.
\end{displaymath}
In particular one has
\begin{align*}
  h_{\psi^{\mathbf{x}}}
         (\mathbf{x})&=\left (\frac{1}{2\pi
     \sigma^2}\right)^{\frac{3}{2}}
e^{-\frac{1}{2\sigma^2}\mathbf{x}^2}
&
\textsf{Var}_{i} (h_{\psi^{\mathbf{x}}})&=\sigma^2
&& i=x,y,z
\intertext{and}
  h_{\psi^{\mathbf{p}}}
         (\mathbf{p})&=\left (\frac{2\sigma^2}{\pi
     \hbar^2}\right)^{\frac{3}{2}}e^{-\frac{2\sigma^2}{\hbar^2}\mathbf{p}^2}
&
\textsf{Var}_{i} (h_{\psi^{\mathbf{p}}})&=\frac{\hbar^2}{4\sigma^2}
&& i=x,y,z
\end{align*}
so that
\begin{displaymath}
   \textsf{Var}_{i} (h_{\psi^{\mathbf{x}}})\textsf{Var}_{i} (h_{\psi^{\mathbf{p}}})= \frac{\hbar^2}{4}
\qquad i=x,y,z.
\end{displaymath}
The POVM now reads
\begin{equation}
\label{exp}
       F^{\mathbf{x},\mathbf{p}}(M\times N)=\frac{1}{(2\pi\hbar)^3}\int_{M}
      d^3\!\mathbf{x}_0\int_{N}d^3\!\mathbf{p}_0
  \,
  |\psi_{\mathbf{x}_{0}\mathbf{p}_{0}}\rangle\langle
  \psi_{\mathbf{x}_{0}\mathbf{p}_{0}} |
\end{equation}
with marginals
\begin{displaymath}
   F^{\mathbf{x}} (M)=\int_{M}d^3\!\mathbf{x}_{0}\int_{\mathbb{R}^3}
   d^3\!\mathbf{x}\,  
\left (\frac{1}{2\pi\sigma^2}\right)^{\frac{3}{2}}e^{-\frac{1}{2\sigma^2}(\mathbf{x}-\mathbf{x}_0)^2}
|\mathbf{x}\rangle
\langle\mathbf{x}|
\end{displaymath}
and
\begin{displaymath}
   F^{\mathbf{p}} (N)=\int_{N}d^3\!\mathbf{p}_{0}\int_{\mathbb{R}^3}
   d^3\!\mathbf{p}\,  
\left (\frac{2\sigma^2}{\pi
     \hbar^2}\right)^{\frac{3}{2}}e^{-\frac{2\sigma^2}{\hbar^2}(\mathbf{p}-\mathbf{p}_0)^2}
|\mathbf{p}\rangle
\langle\mathbf{p}|
\end{displaymath}
for position and momentum respectively. It is now clear that depending
on the value of $\sigma$ one can have more or less coarse-grained
position and momentum observables. No limit on $\sigma$ can however be
taken in order to have a sharp observable for both position and
momentum. In the limit $\sigma \rightarrow 0$ one has as before
$F^{\mathbf{x}}\rightarrow E^{\mathbf{x}}$, but the marginal for
momentum would identically vanish, intuitively corresponding to a
complete lack of information on momentum, and vice versa.

As a last remark we would like to stress that despite the fact that
PVM are only a very particular case of POVM, corresponding to most
accurate measurements, self-adjoint operators, which are in one to one
correspondence to PVM, do play a distinguished and very important
role in quantum mechanics as generators of symmetry transformations,
according to Stone's theorem. 

\subsection{Measurements and state transformations}
\label{sec:meas-state-transf}

The formalism presented up to now allows to describe in the most
general way the state corresponding to a given preparation procedure 
and the statistics of the experimental outcomes obtained by feeding a
certain registration procedure by such a state. It is however also of
interest to have information not only on the statistics of the
outcomes, which amounts to provide a classical probability density,
but also to specify the state obtained as a consequence of such a
measurement, provided the system does not simply get absorbed. This
makes it possible to deal e.g. with repeated consecutive
measurements, allowing for a description of continual measurement in quantum
mechanics\cite{Barchielli2006a}, as well as to use the combination of
initial preparation and registration apparata altogether as a new
preparation apparatus, preparing states according to the value of a
certain observable.

\subsubsection{State transformations with a measuring character as instruments}
\label{sec:state-transf-with}

The mathematical object characterizing a state transformation as a
consequence of a given measurement is called instrument and is
generally given by a mapping defined on the outcome space of the
measurement, e.g. $\mathcal{B}
(\mathbb{R}^3)$ in the examples we have considered, and taking values
in the space of bounded mappings acting on the space of trace class
operators, obeying the following requirements
\begin{align*}
   \mathcal{F} (\cdot):\mathcal{B} (\mathbb{R}^3) &\rightarrow \mathcal{B} (\mathcal{T}(\mathcal{H}))
\\
M &\rightarrow \mathcal{F} (M)
\end{align*}
\begin{flalign*}
& \Tr \mathcal{F}  (\mathbb{R}^3)[\rho]=\Tr \rho
% \\
% &  \mathcal{F}(\emptyset)=0
\\
& \mathcal{F} (\cup_{i} M_i)=\sum_{i}\mathcal{F} (M_i) \qquad M_i\cap M_j=\emptyset \quad
i\not = j
\end{flalign*}
where for each $M\in\mathcal{B}
(\mathbb{R}^3)$ the mapping $\mathcal{F} (M)$ is completely positive and generally
trace decreasing, transforming trace class operators in
trace class operators. $\mathcal{F} (M)$ is often called operation,
and the mapping $\mathcal{F}$ is normalized in the sense that
$\mathcal{F} (\mathbb{R}^3)$ is trace preserving. As for any
completely positive mapping for fixed $M$ one has for  $\mathcal{F}
(M)$ a Kraus representation
\begin{displaymath}
    \mathcal{F} (M)[\rho]=\sum_{i}V_i\rho V_i^{\dagger} .
\end{displaymath}
The trace class operator
\begin{displaymath}
   \mathcal{F} (M)[\rho_{in}]
\end{displaymath}
whose trace is generally less than one, describes the subcollection of
systems obtained by asking the outcome of the measurement to be in $M\in\mathcal{B}
(\mathbb{R}^3)$. After the measurement in fact the transformed system
can be sorted according to the outcome of the measurement itself. The
transformed state according to a measurement without readout,
i.e. without making any selection with respect to the result of the
measurement (the so-called a priori state) is given by
\begin{displaymath}
      \rho_{out} ={\mathcal{F} (\mathbb{R}^3)[\rho_{in}]}
\end{displaymath}
the mapping $\mathcal{F} (\mathbb{R}^3)$ now only describing the
modification on the incoming state $\rho_{in}$ as a consequence of its
interaction with the registration apparatus. The transformed state
conditioned on the result of the measurement, i.e. the new state
obtained by sorting out only the systems for which the outcome of the
measurement was in  $M\in\mathcal{B}
(\mathbb{R}^3)$ (the so-called a posteriori state) is given by
\begin{displaymath}
      \rho_{out} (M) =\frac{\mathcal{F} (M)[\rho_{in}]}{\Tr \mathcal{F} (M)[\rho_{in}]}.
\end{displaymath}
Of course an instrument does not only provide the transformed trace
class operator describing the system after its interaction with the
registration apparatus performing the measurement, but also the
statistics of the outcomes. The probability of an outcome $M\in\mathcal{B}
(\mathbb{R}^3)$ for a measurement described by the instrument
$\mathcal{F}$ on an incoming state $\rho$ is given by the formula
\begin{displaymath}
   \Tr \mathcal{F} (M)[\rho],
\end{displaymath}
which can also be expressed by means of a POVM $F$ uniquely determined
by the instrument $\mathcal{F}$ as follows
\begin{displaymath}
  F (M) =\mathcal{F}' (M)[\openone]
\end{displaymath}
where $\mathcal{F}'$ denotes the adjoint mapping with respect to the
trace evaluation
\begin{displaymath}
   \Tr \mathcal{F} (M)[\rho]=\Tr \openone (\mathcal{F} (M)[\rho])=\Tr (\mathcal{F}' (M)[\openone
      ])\rho=\Tr {F} (M)[\rho].
\end{displaymath}
Note that the correspondence between instruments and POVM is not one
to one. In fact there are different registration apparata, possibly
leading to quite different transformations on the incoming state,
which however all provide a measurement of the same observable. As a
consequence while an instrument uniquely defines the associated POVM
as outlined above, generally infinite different instruments are
compatible with a given POVM, corresponding to different macroscopic
implementation of measurements of the same observable. Note further
that if the mapping $\mathcal{F}$ is reversible it is
necessarily given by a unitary transformation and therefore does not
have any measuring character, the system transforms in a reversible
way because of its free evolution described by a self-adjoint
operator. A very special case of instrument can be obtained starting from
the knowledge of an observable given as self-adjoint operator,
$A=\sum_{i}a_i E_i $, where $\{E_i\}$ is a collection of mutually
orthogonal projection operators, summing up to the identity,
$E_i=E_i^2$ and $\sum_{i}E_i=\openone$. Then the mapping
\begin{displaymath}
   \mathcal{F} (M)[\rho]=\sum_{\{i|a_i\in M\}}E_i \rho E_i
\end{displaymath}
actually is an instrument describing the state transformation of an
incoming state $\rho$ as a consequence of the measurement of the
observable $A$ as predicted by von Neumann's projection postulate. If
the experimenter detects the value $a_i$ for $A$ the  transformed
state is given by
\begin{displaymath}
   \mathcal{F} (\{a_i \})[\rho]=E_i \rho E_i.
\end{displaymath}
This instrument has the peculiar property of being repeatable, in the
sense that from
\begin{displaymath}
    \mathcal{F} (M) [\mathcal{F} (N)[\rho]]=\sum_{\{i|a_i\in M\cap N \}}E_i \rho E_i
\end{displaymath}
follows
\begin{displaymath}
   \mathcal{F} (\{a_i \})[\mathcal{F} (\{a_i \})[\rho]]=\mathcal{F} (\{a_i \})[\rho]=E_i \rho E_i,
\end{displaymath}
that is to say subsequent measurements of the same observable do
always lead to the same result, which implicitly means an absolute
precision in the measurement of the observable. As it appears this is
a very particular situation, which can only be realized for a
measurement in the sense of PVM of an observable with discrete spectrum.

\subsubsection{Example of instrument for position and momentum}
\label{sec:example-instr-posit}

We now provide an example of instrument corresponding to the
description of a state transformation taking place by jointly
measuring position and momentum of a particle in $L^2 (\mathbb{R}^3)$,
whose uniquely associated POVM is just the one given
in~\eqref{exp}. Consider in fact the mapping
\begin{displaymath}
           \mathcal{F}^{\mathbf{x},\mathbf{p}} (M\times N)[\rho]
           =\frac{1}{(2\pi\hbar)^3}\int_{M}
           d^3\!\mathbf{x}_{0}\int_{N}d^3\!\mathbf{p}_{0}
\,
           |\psi_{\mathbf{x}_{0}\mathbf{p}_{0}}\rangle\langle
  \psi_{\mathbf{x}_{0}\mathbf{p}_{0}} |\rho|\psi_{\mathbf{x}_{0}\mathbf{p}_{0}}\rangle\langle
  \psi_{\mathbf{x}_{0}\mathbf{p}_{0}} |
\end{displaymath}
built in terms of the normalized Gaussian wave packets
$|\psi_{\mathbf{x}_{0}\mathbf{p}_{0}}\rangle$ centred in
$(\mathbf{x}_{0},\mathbf{p}_{0})$. This mapping depends on the
interval $M\times N$ as a $\sigma$-additive measure, thanks to the
fact that it is expressed by means of an operator density with respect
to the Lebesgue measure. Normalization is ensured by the completeness
relation for Gaussian wave packets
\begin{displaymath}
   \frac{1}{(2\pi\hbar)^3}\int_{\mathbb{R}^3}
           d^3\!\mathbf{x}_{0}\int_{\mathbb{R}^3}d^3\!\mathbf{p}_{0}
\,
           |\psi_{\mathbf{x}_{0}\mathbf{p}_{0}}\rangle\langle
  \psi_{\mathbf{x}_{0}\mathbf{p}_{0}} |=\openone,
\end{displaymath}
leading to
\begin{align*}
           \Tr \mathcal{F}^{\mathbf{x},\mathbf{p}} (\mathbb{R}^3\times \mathbb{R}^3)[\rho]
%            &=\frac{1}{(2\pi\hbar)^3}\int_{\mathbb{R}^3}
%            d^3\!\mathbf{x}_{0}\int_{\mathbb{R}^3}d^3\!\mathbf{p}_{0}
% \, \Tr
%            |\psi_{\mathbf{x}_{0}\mathbf{p}_{0}}\rangle\langle
%   \psi_{\mathbf{x}_{0}\mathbf{p}_{0}} |\rho|\psi_{\mathbf{x}_{0}\mathbf{p}_{0}}\rangle\langle
%   \psi_{\mathbf{x}_{0}\mathbf{p}_{0}} |
% \\
&=\frac{1}{(2\pi\hbar)^3}\int_{\mathbb{R}^3}
           d^3\!\mathbf{x}_{0}\int_{\mathbb{R}^3}d^3\!\mathbf{p}_{0}
\, \langle
  \psi_{\mathbf{x}_{0}\mathbf{p}_{0}} |\rho|\psi_{\mathbf{x}_{0}\mathbf{p}_{0}}\rangle
\\
&=
\Tr \rho,
\end{align*}
while the adjoint mapping acting on the identity operator
\begin{align*}
           \mathcal{F}^{\mathbf{x},\mathbf{p}}{}'(M\times N)[\openone]
           &=\frac{1}{(2\pi\hbar)^3}\int_{M}
           d^3\!\mathbf{x}_{0}\int_{N}d^3\!\mathbf{p}_{0}
\,
           |\psi_{\mathbf{x}_{0}\mathbf{p}_{0}}\rangle\langle
  \psi_{\mathbf{x}_{0}\mathbf{p}_{0}} |\openone|\psi_{\mathbf{x}_{0}\mathbf{p}_{0}}\rangle\langle
  \psi_{\mathbf{x}_{0}\mathbf{p}_{0}} |
\\
&=\frac{1}{(2\pi\hbar)^3}\int_{M}
           d^3\!\mathbf{x}_{0}\int_{N}d^3\!\mathbf{p}_{0}
\, 
|\psi_{\mathbf{x}_{0}\mathbf{p}_{0}}\rangle\langle 
  \psi_{\mathbf{x}_{0}\mathbf{p}_{0}} |
\\
&=
F^{\mathbf{x},\mathbf{p}} (M\times N)
\end{align*}
immediately gives the joint position and momentum POVM considered
in~\eqref{exp}. More generally one can consider an instrument of the form
\begin{multline*}
           \mathcal{F}^{\mathbf{x},\mathbf{p}} (M\times N)
           =\frac{1}{(2\pi\hbar)^3}\int_{M}
           d^3\!\mathbf{x}_{0}\int_{N}d^3\!\mathbf{p}_{0}
\,
\\
\times
W (\mathbf{x}_0,\mathbf{p}_0)\sqrt{S}W^{\dagger} (\mathbf{x}_0,\mathbf{p}_0)
\rho
W (\mathbf{x}_0,\mathbf{p}_0)\sqrt{S}W^{\dagger} (\mathbf{x}_0,\mathbf{p}_0)
\end{multline*}
which is again well-defined due to the relation
\begin{displaymath}
   \frac{1}{(2\pi\hbar)^3}\int_{\mathbb{R}^3}
           d^3\!\mathbf{x}_{0}\int_{\mathbb{R}^3}d^3\!\mathbf{p}_{0}
\, W (\mathbf{x}_0,\mathbf{p}_0){S}W^{\dagger} (\mathbf{x}_0,\mathbf{p}_0)=\openone
\end{displaymath}
where $S$ is a positive operator given by a statistical operator
invariant under rotations, and whose adjoint mapping applied to the
identity
\begin{align*}
           \mathcal{F}^{\mathbf{x},\mathbf{p}}{}'(M\times N)[\openone]
           &=\frac{1}{(2\pi\hbar)^3}\int_{M}
           d^3\!\mathbf{x}_{0}\int_{N}d^3\!\mathbf{p}_{0}
\, W (\mathbf{x}_0,\mathbf{p}_0)S W^{\dagger} (\mathbf{x}_0,\mathbf{p}_0)
\\
&=F^{\mathbf{x},\mathbf{p}} (M\times N)
\end{align*}
coincides with the general expression for a covariant position and
momentum POVM observable given in \eqref{fxp}.

\section{Open systems and covariance}
\label{sec:open-syst-covar}

In the previous Section we have outlined the modern formulation of
quantum mechanics, understood as a probability theory necessary for
the description of the outcomes of statistical experiments involving microscopic systems. In this framework pure states are
generally replaced by statistical operators, observables in the sense
of self-adjoint operators are substituted by mappings taking values in
an operator space, von Neumann's projection postulate is a very
special case of mappings with a measuring character describing state
transformations as a consequence of measurement. As it appears the
notion of mapping taking values in an operator space becomes very
natural and of great relevance. The knowledge of an instrument for
example allows not only to predict the statistics of the outcomes of a
certain measurement, but also the transformation of the state due to
the interaction with the measuring apparatus. On the contrary the
spontaneous transformation of the state of a closed system due to
passing time is described by a very special kind of mappings, unitary
time evolutions uniquely determined by fixing a self-adjoint operator.
Considering more generally an open system, that is to say a system
interacting with some other external system, its irreversible
evolution in time as a consequence of this interaction is determined
by a suitable mapping, whose characterization is a very intricate and
interesting subject, together with the possibility to understand and
describe such an evolution as a measurement effected on the system.
While it is relatively easy to state the general properties that
should be obeyed by such mappings in order to provide a well-defined
time evolution, the characterization of the structure of such mappings
in its full generality is an overwhelmingly complicated problem.
Important and quite general results can however be obtained
considering constraints coming from physical or mathematical
considerations. Also in this Section we will aim at a brief
introduction of key concepts, skipping all mathematical details,
referring the reader
to\cite{Alicki1987a,Alicki2001a,Holevo2001a,Breuer2007,Alicki2002b}
for a more exhaustive presentation.

\subsection{Constraints on dynamical mappings}
\label{sec:constr-dynam-mapp}

In Sect.\ref{sec:quantum-mechanics-as} we have already mentioned two
important constraints applying to mappings describing how a state
transforms in time, both as a consequence of a measurement or of its
dynamical evolution. Actually the two situations are not
of a completely different nature, even though in the first case the
time extension of the interaction between system ad measuring
apparatus is typically though not necessarily assumed to be very short
and neglected, so that the whole transformation is considered as a one
step process. In the case of the evolution of a closed or open
system on the contrary the explicit time dependence of the mapping is essential, while a
decomposition of the mapping according to the measurement outcome for
a certain observable is generally not available.

\subsubsection{Complete positivity}
\label{sec:complete-positivity}

The first constraint was the by now well-known requirement of complete
positivity of the mapping. It is a mathematical condition, which at
the beginning was somewhat mistrusted by physicists, naturally coming
in the foreground in quantum mechanics because of the tensor product
structure of the space in which to describe composite systems, thus
playing an important role in the theory of entanglement. A completely
positive mapping is a mapping which remains positive when extended in
a trivial way, i.e. by taking the tensor product with the identity
mapping, on a composite Hilbert space. Considering a positive mapping
$\mathcal{M}$ in the Schr\"odinger picture, acting on the space of
states $\mathcal{T}(\mathcal{H})$
\begin{align*}
\mathcal{M}: \mathcal{T}(\mathcal{H})
&\rightarrow
\mathcal{T}(\mathcal{H})
\\
 \rho 
&\rightarrow 
\mathcal{M}[\rho]
\end{align*}
complete positivity amounts to the requirement that the mapping
\begin{align*}
\mathcal{M}_n: \mathcal{T}(\mathcal{H}\otimes \mathbb{C}^n)
&\rightarrow
\mathcal{T}(\mathcal{H}\otimes \mathbb{C}^n)
\\
 \rho \otimes \sigma_n
&\rightarrow 
\mathcal{M}[\rho] \otimes \sigma_n
\end{align*}
is positive for any $n\in\mathbb{N}$, with $\sigma_n$ a statistical
operator in $\mathbb{C}^n$. An equivalent requirement can be
formulated on the adjoint mapping $\mathcal{M}'$ in Heisenberg picture
acting on the space of observables $\mathcal{B}(\mathcal{H})$
\begin{align*}
\mathcal{M}': \mathcal{B}(\mathcal{H})
&\rightarrow
\mathcal{B}(\mathcal{H})
\\
B&\rightarrow 
\mathcal{M}'[B]
\end{align*}
$\mathcal{M}'$ is completely positive provided
\begin{align*}
\sum_{i,j=1}^n
\langle \psi_i |  {\cal M}' (  B_i^\dagger   B_j) \psi_j\rangle
\geq 0
\qquad
\{\psi_i\}\subset \mathcal{H},
\qquad
 \{  B_i\}\subset \mathcal{B}(\mathcal{H})
\end{align*}
for any $n\in \mathbb{N}$.
As it was shown by Kraus any completely positive mapping can be
expressed as follows
\begin{displaymath}
   \mathcal{M}[\rho]=\sum_{i}V_i\rho V_i^{\dagger}
\end{displaymath}
with a suitable collection $\{  V_i\}$ of operators also called Kraus
operators, as already mentioned in Sec.\ref{sec:state-transf-with}. As
it appears from this fundamental result, of great significance in
applications, the condition of complete positivity is quite
restrictive, thus allowing for important characterizations.

\subsubsection{Covariance}
\label{sec:covariance}

Another important constraint, this time however only arising in the
presence of symmetries, is given by the requirement of covariance,
already considered in Sec.\ref{sec:covariant-mapping}. For the case of
a mapping defined on an operator space, e.g. in Schr\"odinger
picture sending statistical operators in statistical operators, given
a unitary representation $U (g)$ of the group $G$ on $\mathcal{H}$
the requirement of covariance can be
expressed as follows
\begin{equation}
   \label{covm}
   \mathcal{M}[U (g)\rho U^{\dagger} (g)]=U
   (g)\mathcal{M}[\rho]U^{\dagger} (g)
\qquad
\forall g\in G.
\end{equation}
This condition expresses the fact that the action of the mapping and of
the representation of $G$ on $\mathcal{T}(\mathcal{H})$
commute, and automatically implies the same property for the adjoint mapping
$\mathcal{M}'$ acting on the space of observables
$\mathcal{B}(\mathcal{H})$. Such a condition typically applies when a
symmetry, available in the system one is studying, is not spoiled by
the transformations brought about on the system by letting it interact
with another system, be it a reservoir or a measuring apparatus. The
possibility of giving a general solution of the covariance
equation~\eqref{covm} obviously depends on the unitary representation
of the group and on the class of mappings considered, possibly giving
very detailed information on the general structure of such mappings.

\subsubsection{Semigroup evolution}
\label{sec:semigroup-evolution}

For the case of a closed system we know that the mapping giving the
reversible time evolution is a one parameter group of unitary
transformations, fixed by a self-adjoint Hamiltonian according to
Stone's theorem. A broader class of time evolutions allowing for an
irreversible dynamics can be obtained by relaxing the group property
to a semigroup composition law, corresponding to the existence of a
preferred time direction. In particular one can introduce a so-called
quantum-dynamical semigroup, which is a collection of one-parameter
mappings $\{\mathcal{U}_t\}_{t\in \mathbb{R}_{+}}$ such that
\begin{align*}
\mathcal{U}_t: \mathcal{T}(\mathcal{H})
&\rightarrow
\mathcal{T}(\mathcal{H})
\\
 \rho 
&\rightarrow 
\mathcal{U}_t[\rho]
\end{align*}
is completely positive and trace preserving for any $t\geq 0$, for
$t=0$ one has the identity mapping, and the following semigroup
composition law applies
\begin{equation}
   \label{m}
   \mathcal{U}_t=\mathcal{U}_{t-s}\mathcal{U}_{s}\qquad \forall t\geq
   s\geq 0.
\end{equation}
The semigroup condition \eqref{m} is sometimes called Markov
condition, because it expresses the fact that the time evolution of
the system does not exhibit memory effects, in analogy with the notion
of Markov semigroup in classical probability theory. It tells us that
the evolution up to time $t$ can be obtained by arbitrarily composing
the evolution mapping up to an intermediate time $s$ with an evolution
mapping depending only on the residual time $t-s$ acting on the state
$\rho_s=\mathcal{U}_{s}[\rho_0]$, not referring to the knowledge of
the state at previous times $\{\rho_{t'}\}_{0\leq t' \leq s}$. The
requirement of complete positivity for this family of mappings allows
for a most important characterization of the so-called generator
$\mathcal{L}$ of the quantum-dynamical semigroup, which is the mapping giving the infinitesimal time
evolution, defined through the relation
\begin{displaymath}
   \mathcal{U}_t=e^{t\mathcal{L}}.
\end{displaymath}
According to a celebrated result of Gorini, Kossakowski, Sudarshan and
Lindblad\cite{Gorini1976a,Lindblad1976a} of enormous relevance in the
applications, the general structure of the generator of a
quantum-dynamical semigroup is given in the Schr\"odinger picture by
\begin{equation}
   \label{l}
\mathcal{L}[\rho]=-\frac{i}{\hbar}[H,\rho]+\sum_{j}[L_{j}\rho
      L^{\dagger}_{j}-\frac{1}{2}\{L^{\dagger}_{j}L_{j},\rho\}],
\end{equation}
where $H$ is a self-adjoint operator, and the operators $L_{j}$ are
often called Lindblad operators. The expression~\eqref{l} is also
called a master equation, since it provides the infinitesimal time
evolution of the statistical operator, according to
${d\rho}/{dt}=\mathcal{L}[\rho]$. The key point is now obviously to
determine the explicit expression of $H$ and $L_{j}$ relevant for the
reduced dynamics of the physical system of interest, typically
depending on the external reservoir and the details of the interaction
mechanism. As we shall see further restrictions on $\mathcal{L}$ can
arise as a consequence of an available symmetry not destroyed by the
interaction. Note that introducing the operator
\begin{displaymath}
   K=\frac{i}{\hbar}H +\frac{1}{2}\sum_{j} L^{\dagger}_{j}L_{j}
\end{displaymath}
where the effective Hamiltonian $H$ appears together with an operator
given by $\frac{\hbar}{2}\sum_{j} L^{\dagger}_{j}L_{j}$ which can be formally
seen as an imaginary, optical effective potential, the Lindblad
structure~\eqref{l} can also be written as
\begin{displaymath}
   \mathcal{L}[\rho]=-K \rho -\rho K^{\dagger}
+\sum_{j}L_{j}\rho
      L^{\dagger}_{j},
\end{displaymath}
leading to a Dyson expansion of the semigroup evolution
\begin{multline}
   \label{Dyson}
      \mathcal{U}_t[\rho]=e^{t\mathcal{L}}[\rho]=\mathcal{K}_t[\rho]
\\
+ \sum_{n=1}^{\infty}
      \int \ldots \int_{0\leq t_1\leq \ldots \leq t_n\leq t}
      dt_1\ldots dt_n \,
      \mathcal{K}_{t_1}[\mathcal{J}[\mathcal{K}_{t_2-t_1}\ldots\mathcal{J}[\mathcal{K}_{t-t_n}[\rho]]\ldots]]
   \end{multline}
where the superoperators
\begin{displaymath}
      \mathcal{J}[\rho]=\sum_{j}L_{j}\rho L^{\dagger}_{j}
      \qquad
      \mathcal{K}_t[\rho]=e^{-Kt}
      \rho
      e^{-K^{\dagger}t}
   \end{displaymath}
appear. The formal solution of the time evolution~\eqref{Dyson} has a
quite intuitive physical meaning, in fact it can be seen as a sequence
of relaxing evolutions over a time interval $t$ given by the
contraction semigroup $ \mathcal{K}_t$ interrupted by jumps described
by the completely positive mapping $\mathcal{J}$. The sum over all
possible such evolutions gives the final state. Most recently a
general characterization has been obtained also for a class of
non-Markovian time evolutions which provides a kind of generalization
of the Lindblad result~\eqref{l}, in the sense that the state evolved
up to a given time can be expressed as a mixture of subcollections
(that is positive trace class operators with trace less than one) each
obeying a Lindblad type of master equation, however with different
Lindblad operators\cite{Budini2005a,Breuer2006a,Budini2006a,Breuer2007a,Breuer2007b}, so that the overall
time evolution does no more obey a semigroup composition law as
in~\eqref{m}.

\subsection{Shift-covariance and damped harmonic oscillator}
\label{sec:shift-covar-damp}

As a first example of master equation corresponding to a completely
positive quantum-dynamical semigroup let us consider the well-known
master equation for a damped harmonic oscillator\cite{Englert2002b}, describing e.g. the
damping of an electromagnetic field mode in a cavity. The Hilbert
space of the single mode is given by $\mathcal{H}=l^{2}
(\mathbb{C})$, the square summable sequences over the complex field,
with basis $\{|n\rangle\}$, the ket $|n\rangle$ denoting as usual the eigenvector
with eigenvalue $n\in\mathbb{N}$ of the number operator
$N=a^{\dagger}a$, $a^{\dagger}$ and $a$ being respectively creation
and annihilation operators of a photon in the given mode of frequency
$\omega$. The master equation then reads
\begin{multline}
\label{dho}
   \mathcal{L}_{\scriptscriptstyle DHO}[\rho]=-\frac{i}{\hbar}[H_0
   (N),\rho]
+\eta (N_{\beta} (\omega)+1)
   [ a\rho a^{\dagger}-\frac{1}{2}\{a^{\dagger}a,\rho \} ]
\\
+\eta N_{\beta} (\omega) 
[a^{\dagger}\rho a-\frac{1}{2}\{aa^{\dagger},\rho \} ],
\end{multline}
where $N_{\beta} (\omega)$ denotes the average of the photon number
operator over a thermal distribution
\begin{displaymath}
   N_{\beta} (\omega)=\frac{1}{e^{\beta\hbar\omega}-1}=\frac{1}{2}[\coth({\beta\hbar\omega}/{2})-1 ],
\end{displaymath}
$H_0(N)=\hbar\omega N$ is the free Hamiltonian and $\eta$ the
relaxation rate.

\subsubsection{Dissipation and decoherence for the damped harmonic oscillator}
\label{sec:diss-decoh-damp}

As it is well known such a master equation describes both classical
dissipative effects as well as quantum decoherence effects. To see
this let us first focus on dissipative effects, considering the time
evolution of mean amplitude and mean number of quanta in the
mode. Considering the adjoint mapping of~\eqref{dho}, giving the time
evolution in Heisenberg picture
\begin{multline*}
   \mathcal{L}'_{\scriptscriptstyle DHO}[X]=+\frac{i}{\hbar}[H_0
   (N),\rho]+\eta (N_{\beta} (\omega)+1)
[a^{\dagger}\rho a-\frac{1}{2}\{a^{\dagger}a,\rho \} ]
\\+\eta N_{\beta} (\omega)
[a\rho a^{\dagger}-\frac{1}{2}\{aa^{\dagger},\rho \} ]   ,
\end{multline*}
one can solve the Heisenberg equations of motion for $X\rightarrow a $
and $X\rightarrow N=a^{\dagger}a$, finally obtaining
\begin{align*}
\langle a (t)\rangle&=\Tr (a (t)\rho)=\langle a \rangle e^{-i\omega t -{\frac{\eta}{2}}t}
\\
  \langle N(t) \rangle&=\Tr (N (t)\rho)=\langle N
  \rangle e^{-{\eta}t}+
N_{\beta}
   (\omega)(1 -e^{-{\eta}t}),
\end{align*}
where $a (t)$, $N (t)$ denote Heisenberg operators at time $t$, with
$a=a (0)$, $N=N (0)$ the corresponding Schr\"odinger operators. The
mean amplitude of the mode thus rotates in the complex plane, vanishing
for long enough times, with a decay rate given by $(\eta/2)^{-1}$, the
population of the mode goes from the initial value to a final thermal
distribution with a decay rate given by
$\eta^{-1}$. For the study of decoherence we shall consider the
time evolution of an initial state given by a coherent superposition
of two coherent states characterized by two amplitudes $\alpha$ and
$\beta$\cite{Breuer2007}. Setting
\begin{displaymath}
   \rho_0=\frac{1}{\mathcal{N}_0}
[|\alpha\rangle\langle\alpha|+|\beta\rangle\langle\beta|+|\alpha\rangle\langle\beta|+h.c.]
\end{displaymath}
one can look for the time evolved state, exploiting the fact that
coherent states remain coherent states under time evolution. Working
for simplicity at zero temperature, so that $N_{\beta} (\omega)=0$ one
has
\begin{multline*}
   \rho_{t}= \frac{1}{\mathcal{N}_t}
   [|\alpha
   (t)\rangle\langle\alpha(t)|+|\beta(t)\rangle\langle\beta(t)|
   \\
   +e^{-\frac{1}{2}|\alpha-\beta|^2
     (1-e^{-\eta t})}|\alpha(t)\rangle\langle\beta(t)|+h.c.]
\end{multline*}
where $\alpha(t)=\alpha e^{-i\omega t -{\frac{\eta}{2}}t}$ and similarly for
$\beta (t)$. As it immediately appears the so-called coherences, that
is to say the off-diagonal matrix elements of the statistical
operator, are suppressed with respect to diagonal ones by a factor
that for long enough times is given by
\begin{displaymath}
   e^{-\frac{1}{2}|\alpha-\beta|^2}=|\langle\alpha|\beta\rangle|
\end{displaymath}
that is to say the modulus of the overlap of the two coherent states, a
tiny  quantity for two macroscopically distinguishable states of the
electromagnetic field.

\subsubsection{Structure of the mapping and covariance}
\label{sec:struct-mapp-covar}

Is is immediately apparent that the master equation~\eqref{dho} is an
example of realization of the Lindblad structure~\eqref{l} with just
two Lindblad operators given by
\begin{displaymath}
   L_1=\sqrt{\eta (N_{\beta} (\omega)+1)}a
\qquad
   L_2=\sqrt{\eta N_{\beta} (\omega)}a^{\dagger},
\end{displaymath}
with Hamiltonian
\begin{displaymath}
   H_0(N)=\hbar\omega N
\end{displaymath}
and stationary solution
\begin{equation}
\label{sdho}
     w\propto e^{-\beta H_0(N)}.
  \end{equation}
More than this it is also covariant under the action of the group $U
(1)$. Consider in fact the unitary representation of $U (1)$ on $\mathcal{H}=l^{2}
(\mathbb{C})$ given by
\begin{displaymath}
   U (\theta)=e^{i\theta N} \qquad \theta\in[0,2\pi],
\end{displaymath}
where $N$ is the usual number operator. The
master equation~\eqref{dho} is covariant under this unitary
representation of the group $U(1)$ according to
\begin{equation}
\label{u1}
   \mathcal{L}_{\scriptscriptstyle DHO}[U (\theta)\rho U^{\dagger} (\theta)]=U (\theta)\mathcal{L}_{\scriptscriptstyle DHO}[\rho]U^{\dagger} (\theta)
\end{equation}
as can immediately be checked. One can also show that this is the
unique structure of master equation bilinear in $a$ and $a^{\dagger}$
complying with the Lindblad structure, covariant under  $U(1)$ and
admitting~\eqref{sdho} as a stationary state\cite{Vacchini2002b}.

It is also possible to give a complete characterization of the
structure of the generator of a quantum-dynamical semigroup covariant
with respect to  $U(1)$ as in~\eqref{u1}. The general result has been
obtained by Holevo\cite{Holevo1995b,Holevo1993b} and is given by the following expression
\begin{multline*}
\mathcal{L}[\rho]
=
-{i \over \hbar}
        \left[
        H (N)
        ,\rho
        \right]
+
\sum_j
\left[
\vphantom{       -
        \frac 12
        \left \{
       A^{\dagger}_{0j}(N)A_{0j}(N),\rho 
        \right \}
}
A_{0j}(N)\rho A^{\dagger}_{0j}(N)
       -
        \frac 12
        \left \{
       A^{\dagger}_{0j}(N)A_{0j}(N),\rho 
        \right \}
\right]
\\
\shoveright{
+
\sum_{m=1}^{\infty}
\sum_j
\left[
\vphantom{       -
        \frac 12
        \left \{
       A^{\dagger}_{mj}(N)A_{mj}(N),\rho 
        \right \}
}
W^{m}A_{mj}(N)\rho A^{\dagger}_{mj}(N)W^{\dagger}{}^{m}
       -
        \frac 12
        \left \{
       A^{\dagger}_{mj}(N)A_{mj}(N),\rho 
        \right \}
\right]
}
\\
+
\sum_{m=1}^{\infty}
\sum_j
\left[
\vphantom{       -
        \frac 12
        \left \{
       A^{\dagger}_{mj}(N)A_{mj}(N),\rho 
        \right \}
}
W^{\dagger}{}^{m}A_{-mj}(N)\rho A^{\dagger}_{-mj}(N)W^{m}
       -
        \frac 12
        \left \{
       A^{\dagger}_{-mj}(N)P_mA_{-mj}(N),\rho 
        \right \}
\right]
\end{multline*}
where $A_{mj}(N)$ are functions of the number operator $N$, the
generator of the symmetry, the operator $W$ is given by
\begin{displaymath}
   W=\sum_{n=0}^{\infty} |n+1\rangle\langle n|
\end{displaymath}
acting as a shift $|n\rangle \rightarrow |n+1\rangle$ on the basis of
eigenvectors of the number operator, so that this kind of symmetry is also called shift-covariance, while
$P_m$ is the projection on the subspace spanned by
$\{|n\rangle\}_{n=m,\ldots,+\infty}$ given by
\begin{displaymath}
   P_m\equiv \sum_{n=m}^{\infty} |n\rangle\langle n|=W^{m}W^{\dagger}{}^{m}
\end{displaymath}
and one further has
\begin{equation}
\label{weylg}
   U (\theta)W^{m}=e^{i\theta m}W^{m}U
(\theta),
\end{equation}
which compared to~\eqref{weyl} can be seen as a generalized Weyl
relation, expressed by means of the isometric but not unitary
operators $W^{m}$\cite{Holevo1993a}.
Examples of realizations of this general shift-covariant expression
are given by the master equation for the damped harmonic oscillator as
indicated above, corresponding to the choice
\begin{gather*}
   A_1 (N) = \sqrt{\eta N_{\beta} (\omega)}\sqrt{N+1}
\qquad
A_{-1} (N) =  e^{\frac{1}{2}\beta\hbar\omega}\sqrt{\eta N_{\beta}
  (\omega)}\sqrt{N}
\\
A_m (n) = 0 \  m=0, |m|>1
\qquad
H (N) =\hbar\omega N,
\end{gather*}
as can immediately be checked exploiting the polar representation for
the creation and annihilation operators
\begin{displaymath}
   a=W^{\dagger}\sqrt{N}\qquad a^{\dagger}=W\sqrt{N+1}.
\end{displaymath}
A more general structure still preserving the stationary
solution~\eqref{sdho} is given by the choice
\begin{gather*}
A_0 (n) = \eta_0
\qquad   A_m (N) = \sqrt{\eta_m}{ N_{\beta}^{\frac{m}{2}} (\omega)}\sqrt{\frac{(N+m)!}{N!}}
\\
   A_{-m} (N) = e^{\frac{m}{2}\beta\hbar\omega}\sqrt{\eta_m}{ N_{\beta}^{\frac{m}{2}} (\omega)}\sqrt{\frac{N!}{(N-m)!}}
\qquad
H (N) =\hbar\omega N,
\end{gather*}
corresponding to
\begin{multline*}
   \mathcal{L}[\rho]=-\frac{i}{\hbar}[H_0
   (N),\rho]-\eta_0[N,[N,\rho]]
\\
\shoveright{
+
\sum_{m=1}^{+\infty}
\eta_m (N_{\beta} (\omega)+1)^{m}
   [ a^{m}\rho
   a^{\dagger}{}^{m}-\frac{1}{2}\{a^{\dagger}{}^{m}a^{m},\rho \} ]
}
\\
+\sum_{m=1}^{+\infty}
\eta_m N_{\beta}^{m} (\omega) 
[a^{\dagger}{}^{m}\rho a^{m}-\frac{1}{2}\{a^{m}a^{\dagger}{}^{m},\rho \} ]
\end{multline*}
where a phase damping term given by a double commutator with the
number operator, as well as many photon processes with different decay
rates appear.

 \subsection{Rotation-covariance and two-level system}
 \label{sec:rotat-covar-two}

Another example of well-known master equation which can be
characterized in terms of covariance properties comes from the
description of a two-level system interacting with a thermal
reservoir, e.g. a two-level atom in the presence of the radiation
field or a spin in a magnetic field, so that the Hilbert space is now
simply $\mathbb{C}^2$. It corresponds to the so-called Bloch equation
and is typically used in quantum optics and magnetic resonance
theory. Focussing on a two-level atom with transition frequency
$\omega$ and spontaneous emission rate $\eta$ interacting with the
quantized electromagnetic field one has
\begin{multline}
   \label{2ls}
   \mathcal{L}_{\scriptscriptstyle 2LS}[\rho]=-\frac{i}{\hbar}[H_0
   (\sigma_z),\rho]
+\eta (N_{\beta} (\omega)+1)
   [ \sigma_{-}\rho \sigma_{+}-\frac{1}{2}\{\sigma_{+}\sigma_{-},\rho \} ]
\\
+\eta N_{\beta} (\omega) 
[\sigma_{+}\rho \sigma_{-}-\frac{1}{2}\{\sigma_{-}\sigma_{+},\rho \} ] ,
\end{multline}
where $N_{\beta} (\omega)$ is the thermal photon number at the
transition frequency and as usual $\sigma_{\pm}=\frac{1}{2}
(\sigma_x\pm i\sigma_y)$ with $\{\sigma_i\}_{i=x,y,z}$ the Pauli
matrices. 

 \subsubsection{Dissipation and decoherence for the two-level system}
 \label{sec:diss-decoh-two}

As it is well-known this master equation predicts relaxation to a
stationary state which is diagonal in the basis of eigenvectors of the
free Hamiltonian, with a relative population between ground and
excited state determined by the temperature of the bath. This can be
immediately seen considering as usual the adjoint mapping of~\eqref{2ls}
\begin{align*}
      \mathcal{L}_{\scriptscriptstyle 2LS}'[X]=+\frac{i}{\hbar}[H_0
   (\sigma_z),X]
+\eta (N_{\beta} (\omega)+1)
   [ \sigma_{+}X \sigma_{-}-\frac{1}{2}\{\sigma_{+}\sigma_{-},X \} ]
\\
+\eta N_{\beta} (\omega) 
[\sigma_{-}X \sigma_{+}-\frac{1}{2}\{\sigma_{-}\sigma_{+},X \} ] 
\end{align*}
and solve the Heisenberg equations of motion for the operator
$X\rightarrow P_e=| 1\rangle\langle 1|$ representing  the population in
the excited state
\begin{displaymath}
   \langle P_e(t) \rangle=\Tr (P_e (t)\rho)=\langle P_e
  \rangle e^{-\bar{\eta}t}+
\frac{N_{\beta}
   (\omega)}{2N_{\beta}
   (\omega)+1}(1 -e^{-\bar{\eta}t}),
\end{displaymath}
which also fixes $\langle P_g(t) \rangle=1-\langle P_e(t) \rangle$ due
to the normalization condition, where we have denoted as $\bar{\eta}=\eta ({2N_{\beta}
   (\omega)+1})$ the total transition rate and $P_e=P_e (0)$. Due to
 the simplicity of the equation also the evolution in time of the
 coherences is easily determined, now considering the evolution in
 Heisenberg picture of the operator $X\rightarrow
 C=|0\rangle\langle1|$, which is given by
 \begin{displaymath}
    \langle C (t)\rangle=\Tr (C (t)\rho)=\langle C \rangle e^{-i\omega t -\frac{\bar{\eta}}{2}t}
 \end{displaymath}
where again $C=C (0)$. With elapsing time the populations reach a
stationary value, while coherences get suppressed.

 \subsubsection{Structure of the mapping and covariance}
 \label{sec:struct-mapp-covar-1}

In strict analogy with the results presented in
Sect.\ref{sec:struct-mapp-covar} also the master equation~\eqref{2ls}
can be immediately recast in Lindblad form~\eqref{l} with two Lindblad
operators given by
\begin{displaymath}
   L_1=\sqrt{\eta (N_{\beta} (\omega)+1)}\sigma_{-}
\qquad
   L_2=\sqrt{\eta N_{\beta} (\omega)}\sigma_{+}
\end{displaymath}
and Hamiltonian
\begin{displaymath}
   H_0(\sigma_z)=\frac{\hbar\omega}{2} \sigma_z
\end{displaymath}
admitting the stationary solution
\begin{equation}
\label{s2ls}
     w\propto e^{-\beta H_0(\sigma_z)}.
  \end{equation}
The master equation is also covariant under the group $SO (2)$,
describing rotations along a given axis and
isomorphic to $U (1)$,
which accounts for the similarity between the two equations, even
though important differences appear due to the different Hilbert spaces
in which the two systems are described and the group represented. Consider in fact the representation of $SO (2)$ in  $\mathbb{C}^2$
\begin{displaymath}
   U (\phi)=e^{\frac{i}{\hbar}\phi S_z}\qquad \phi\in[0,2\pi]
\end{displaymath}
with $S_z=\frac{\hbar}{2}\sigma_z$: one immediately checks
that~\eqref{2ls} is covariant in the sense that
\begin{equation}
\label{so2}
   \mathcal{L}_{\scriptscriptstyle 2LS}[U (\phi)\rho U^{\dagger} (\phi)]=U (\phi)\mathcal{L}_{\scriptscriptstyle 2LS}[\rho]U^{\dagger} (\phi).
\end{equation}
Also in this case it is possible to provide the explicit expression of
the generator of a rotation-covariant quantum-dynamical semigroup in
the sense of~\eqref{so2}. It takes the simple form\cite{Artemev1989}
\begin{displaymath}
   \mathcal{L}[\rho]=-\frac{i}{\hbar}[H
   (\sigma_z),\rho]
+
\sum_{m=0,\pm 1}c_m [T_{1m}\rho T_{1m}^{\dagger}-\frac{1}{2}\{T_{1m}^{\dagger}T_{1m},\rho \} ]
\end{displaymath}
where the $c_m$ are positive constants and $T_{1m}$ are irreducible
tensor operators given by
\begin{displaymath}
   T_{11}=-\frac{1}{\sqrt{2}} (\sigma_x + i \sigma_y) \qquad
   T_{10}=\sigma_z\qquad T_{1-1}=\frac{1}{\sqrt{2}} (\sigma_x - i \sigma_y) 
\end{displaymath}
or equivalently in terms of $\sigma_z$, $\sigma_{+}$ and $\sigma_{-}$
in order to allow for a direct comparison with~\eqref{2ls}
\begin{multline*}
   \mathcal{L}[\rho]=-\frac{i}{\hbar}[H
   (\sigma_z),\rho]
-\frac{c_0}{2}[\sigma_z,[\sigma_z,\rho]]
\\
+2 c_{-1}[ \sigma_{-}\rho \sigma_{+}-\frac{1}{2}\{\sigma_{+}\sigma_{-},\rho \} ]
+2 c_1 [\sigma_{+}\rho
\sigma_{-}-\frac{1}{2}\{\sigma_{-}\sigma_{+},\rho \} ].
\end{multline*}

The use of the word rotation-covariance hints at the fact that such a
master equation applies e.g. to a spin $1/2$ in an environment with
axial symmetry, so that one has invariance under rotations along a given
axis. More generally the full rotation group $SO (3)$ can be
considered, and a general characterization also exists for the structure of generators
of semigroups acting on $\mathbb{C}^n$ and invariant under $SO (3)$\cite{Gorini1978a}, relevant for the
description of relaxation of a spin $j$ under the influence of
isotropic surroundings.

\subsection{Translation-covariance and quantum Brownian motion}
\label{sec:transl-covar-quant}

As a further example of the concepts introduced above we will consider
the master equation for the description of quantum Brownian motion,
which applies to the motion of a massive test particle in a gas of
lighter particles. The Hilbert space of relevance is here given by
$L^2 (\mathbb{R}^3)$, and denoting the usual position and momentum
operators as $\hat{\mathbf{x}}$ and $\hat{\mathbf{p}}$ the general structure of
the master equation reads
\begin{multline}
   \label{qbm}
   \mathcal{L}_{\scriptscriptstyle QBM}[\rho]=-\frac{i}{\hbar}\left[H_0 (\hat{\mathbf{p}}),\rho\right]
-\frac{i}{\hbar}\frac{\eta}{2}\sum_{i} [\hat{x}_i,\{\hat{p}_i,\rho\}]
\\
-\frac{D_{pp}}{\hbar^2}\sum_{i} [\hat{x}_i,[\hat{x}_i,\rho]]
-\frac{D_{xx}}{\hbar^2}\sum_{i} [\hat{p}_i,[\hat{p}_i,\rho]],
\end{multline}
where we have assumed isotropy for simplicity, the index $i=x,y,z$
denoting the different Cartesian coordinates and $H_0
(\hat{\mathbf{p}})={\hat{\mathbf{p}}^2}/{2M}$ is the free Hamiltonian. This and similar types of
master equation, bilinear in the position and momentum operators, do
appear in different contexts, leading to different microscopic
expressions for the coefficients, as well as to the appearance of
other terms such as a double commutator $[\hat{x}_i,[\hat{p}_i,\rho]]$ with position and momentum 
operators\cite{Sandulescu1987a,Breuer2007}. We will here have in mind
the dynamics of a test particle interacting through collisions with a
homogeneous gas\cite{Vacchini2000a,Oconnell2001a,Vacchini2001c}, so that the coefficients read
\begin{displaymath}
   D_{pp}=\eta\frac{M}{\beta}
\qquad
D_{xx}=\eta\frac{\beta\hbar^2}{16 M}
\end{displaymath}
with $M$ the mass of the test particle. 

\subsubsection{Dissipation and decoherence for quantum Brownian motion}
\label{sec:diss-decoh-quant}

As in the previous case we want to briefly point out how such a
master equation describes both dissipative and decoherence
effects. Regarding dissipation one has similarly to the classical case
that the mean value of momentum is driven to zero, while the average
of the squared momentum goes to the equipartition value fixed by the
gas temperature. The effect of decoherence typically manifests  itself in
the fact that superpositions of spatially
macroscopically distinguished states are quickly suppressed. Let us
first focus on dissipation, considering the adjoint mapping
of~\eqref{qbm} in Heisenberg picture
\begin{multline*}
   \mathcal{L}_{\scriptscriptstyle QBM}'[X]=+\frac{i}{\hbar}\left[H_0 (\hat{\mathbf{p}}),X\right]
+\frac{i}{\hbar}\frac{\eta}{2}\sum_{i} \{\hat{p}_i,[\hat{x}_i,X]\}
\\
\nonumber
-\frac{D_{pp}}{\hbar^2}\sum_{i} [\hat{x}_i,[\hat{x}_i,X]]
-\frac{D_{xx}}{\hbar^2}\sum_{i} [\hat{p}_i,[\hat{p}_i,X]].
\end{multline*}
As already mentioned the observables of interest are given by the
momentum operators $X\rightarrow \hat{\mathbf{p}}$ and the kinetic energy
$X\rightarrow E={\hat{\mathbf{p}}^2}/{2M}$, whose mean values evolve according to
\begin{align*}
\langle \hat{\mathbf{p}} (t)\rangle&=\Tr (\hat{\mathbf{p}} (t)\rho)=\langle \hat{\mathbf{p}} \rangle e^{-{\eta}t}
\\
  \langle E(t) \rangle&=\Tr (E (t)\rho)=\langle E
  \rangle e^{-2{\eta}t}+
\frac{3}{2\beta}
   (1 -e^{-2{\eta}t}),
\end{align*}
where again $\hat{\mathbf{p}} (t)$ and $E (t)$ denote Heisenberg operators at time
$t$. The average momentum thus relaxes to zero with a decay rate
$\eta^{-1}$, while the mean kinetic energy reaches the equipartition
value with a rate $(2\eta)^{-1}$. 
We now concentrate on the study of decoherence, both for position and
momentum. This can be done considering the off-diagonal matrix
elements in both position and momentum of a statistical operator
evolved in time according to~\eqref{qbm}. To do this we
exploit the knowledge of the exact solution\cite{Bassi2005a},
neglecting however the contribution $-\frac{i}{\hbar}\frac{\eta}{2}\sum_{i}
[\hat{x}_i,\{\hat{p}_i,\rho\}]$ responsible for dissipative effects, that  would
lead to too cumbersome an expression. Considering an initial state
$\rho_0$ the state up to time $t$ reads in the momentum representation
\begin{multline*}
   \langle \mathbf{p}|\rho_{t}|\mathbf{q}\rangle
=
e^{- \frac{D_{xx}}{\hbar^2}(\mathbf{p}-\mathbf{q})^2t}
e^{-\frac{1}{12} \frac{D_{pp}}{\hbar^2}\left
     (\frac{(\mathbf{p}-\mathbf{q})}{M} t\right)^2 t}
\\
\times
\left ( {\frac{\hbar^2}{4\pi D_{pp} t}} \right)^{3/2}\int d^3\!\mathbf{k} \,
e^{-\frac{\hbar^2\mathbf{k}^2}{4D_{pp} t}}
\langle \mathbf{p}-\mathbf{k}|\rho_{0}|\mathbf{q}-\mathbf{k}\rangle.
\end{multline*}
It immediately appears that off-diagonal matrix elements are quickly
suppressed with elapsing time, depending on their separation
$(\mathbf{p}-\mathbf{q})^2$. The factor depending on the coefficient
$D_{xx}$ is due to the momentum localization term $\sum_{i}
[\hat{p}_i,[\hat{p}_i,\rho]]$, while the factor where the coefficient $D_{pp}$
appears is due to the position localization term $\sum_{i}
[\hat{x}_i,[\hat{x}_i,\rho]]$, this also suppresses coherences in momentum because
different momentum states quickly lead to spatial separation, so that
the position localization mechanism is again of relevance. As far as
coherences in position are concerned one has to consider the matrix
elements of $\rho_t$ in the position representation. It is here
convenient to express the exact solution $\rho_t$ in terms of the
solution $\rho_{t}^{S}$ of the free Schr\"odinger equation. One has
the quite cumbersome expression
% \begin{multline*}
%   \langle \mathbf{x}|\rho_{t}|\mathbf{y}\rangle
% =
% e^{-\frac{D_{pp}}{\hbar^2} (\mathbf{x}-\mathbf{y})^2 t
% \left[1-\frac{1}{4}\frac{D_{pp}}{M^2}t^2 \frac{1}{[D_{xx}+ \frac{1}{3}\frac{D_{pp}}{M^2}t^2]} \right]
% }
% \left (\frac{\hbar^2}{4\pi[D_{xx}+ \frac{1}{3}\frac{D_{pp}}{M^2}t^2]t}\right)^{3/2}
% \\
% \times
% \int d^3\!\mathbf{z}\,
% e^{-\frac{\hbar^2\mathbf{z}^2}{4[D_{xx}+ \frac{1}{3}\frac{D_{pp}}{M^2}t^2]t}}
% e^{\frac{i}{\hbar}\frac{1}{2}\frac{D_{pp}}{M}
%   \frac{\mathbf{z}\cdot(\mathbf{x}-\mathbf{y})t}{[D_{xx}+ \frac{1}{3}\frac{D_{pp}}{M^2}t^2]}}
% \langle \mathbf{x}-\mathbf{z}|\rho_{t}^{S}|\mathbf{y}-\mathbf{z}\rangle
% \end{multline*}
\begin{multline*}
  \langle \mathbf{x}|\rho_{t}|\mathbf{y}\rangle
=
e^{-\frac{D_{pp}}{\hbar^2} (\mathbf{x}-\mathbf{y})^2 t
\left[1-\frac{D_{pp}}{4M^2}t^2 \frac{1}{\left[D_{xx}+ \frac{D_{pp}}{3M^2}t^2\right]} \right]
}
\left (\frac{\hbar^2}{4\pi\left[D_{xx}+ \frac{D_{pp}}{3M^2}t^2\right]t}\right)^{3/2}
\\
\times
\int d^3\!\mathbf{z}\,
e^{-\frac{\hbar^2\mathbf{z}^2}{4\left[D_{xx}+ \frac{D_{pp}}{3M^2}t^2\right]t}}
e^{\frac{i}{\hbar}\frac{D_{pp}}{2M}
  \frac{\mathbf{z}\cdot(\mathbf{x}-\mathbf{y})t}{\left[D_{xx}+ \frac{D_{pp}}{3M^2}t^2\right]}}
\langle \mathbf{x}-\mathbf{z}|\rho_{t}^{S}|\mathbf{y}-\mathbf{z}\rangle
\end{multline*}
which is essentially given by a convolution of the free solution with
a Gaussian kernel, multiplied by an exponential factor suppressing
off-diagonal matrix elements according to their distance
$(\mathbf{x}-\mathbf{y})^2$ in space. Spatially macroscopically
distant states are again very quickly suppressed.

\subsubsection{Structure of the mapping and covariance}
\label{sec:struct-mapp-covar-2}

The master equation~\eqref{qbm} can also be written manifestly in
Lindblad form, as it can be seen introducing a single Lindblad operator for
each Cartesian direction
\begin{displaymath}
   L_i=\sqrt{\eta}a_i
\end{displaymath}
with
\begin{displaymath}
   a_i=\frac{1}{\sqrt{2}\lambda_{\mathrm{th}}}\left
      (\hat{x}_i+\frac{i}{\hbar}\lambda_{th}^2 \hat{p}_i\right)
\qquad
\lambda_{\mathrm{th}}=\sqrt{\frac{\beta\hbar^2}{4 M}}
\qquad
[a_i,a^{\dagger}_j]=\delta_{ij}
\end{displaymath}
and the effective Hamiltonian
\begin{displaymath}
   H=H_0 (\hat{\mathbf{p}})+\frac{\eta}{2}\sum_{i}\{\hat{x}_i,\hat{p}_i\},
\end{displaymath}
leading to
\begin{displaymath}
   \mathcal{L}_{\scriptscriptstyle QBM}[\rho]=-\frac{i}{\hbar}\left[H_0(\hat{\mathbf{p}})
      +\frac{\eta}{2}\sum_{i}\{\hat{x}_i,\hat{p}_i\},\rho\right]
+\eta\sum_{i}[a_i\rho a_i^{\dagger}-\frac{1}{2}\{a_i^{\dagger}a_i,\rho \} ]
\end{displaymath}
with stationary solution
\begin{equation}
   \label{sqbm}
   w\propto e^{-\beta H_0(p)}.
\end{equation}
Also the quantum Brownian motion master equation is characterized by
covariance under the action of a symmetry group which in this case is
the group $\mathbb{R}^3$ of translations. Given the unitary representation
\begin{equation}
\label{a}
   U (\mathbf{a})=e^{\frac{i}{\hbar}\mathbf{a}\cdot\hat{\mathbf{p}}}\qquad \mathbf{a}\in\mathbb{R}^3
\end{equation}
of the group of translations on $L^2 (\mathbb{R}^3)$ where the
operators $\hat{\mathbf{p}}$ act as generator of the symmetry, one can
indeed immediately check that~\eqref{qbm} is covariant under this
representation in the sense that
\begin{equation}
\label{r}
   \mathcal{L}_{\scriptscriptstyle QBM}[U (\mathbf{a})\rho U^{\dagger} (\mathbf{a})]=U (\mathbf{a})\mathcal{L}_{\scriptscriptstyle QBM}[\rho]U^{\dagger} (\mathbf{a}).
\end{equation}
In this case however at odds with the case of the master equation for
the damping harmonic oscillator the three requirements of Lindblad
structure, covariance under $\mathbb{R}^3$ and stationary state given
by~\eqref{sqbm} do not uniquely fix the form of a master equation at
most bilinear in the operators $\hat{\mathbf{x}}$ and $\hat{\mathbf{p}}$\cite{Vacchini2002b}. 

A most important general characterization of structures of generators
of quantum-dynamical semigroups covariant under translations has been
obtained by Holevo, relying on a non-commutative quantum
generalization\cite{Holevo1993a,Holevo1995a,Holevo1998} of the
classical L\'evy-Khintchine formula (see e.g.\cite{Feller1971,Breuer2007}). In this case
the generator can be written as follows
\begin{displaymath}
\mathcal{L}[\rho]=-\frac{i}{\hbar}
        \left[
        H ( \hat{\mathbf{p}})
        ,\rho 
        \right]+\mathcal{L}_{G}[\rho]+\mathcal{L}_{P}[\rho],
\end{displaymath}
where $H ( \hat{\mathbf{p}})$ is a self-adjoint operator only depending on
the momentum operators, while $\mathcal{L}_{G}$ and $\mathcal{L}_{P}$
correspond to a Gaussian and a Poisson component, as in the
L\'evy-Khintchine formula, and
are given by
\begin{multline}
\label{g}
\mathcal{L}_{G}[\rho]=-\frac{i}{\hbar} \left[\hat{y}_0+\frac{1}{2i}
\sum_{k=1}^{r}
(\hat{y}_k L_k (\hat{\mathbf{p}}) -L_k^{\dagger}
(\hat{\mathbf{p}})\hat{y}_k), \rho \right]
\\
\shoveleft{+\sum_{k=1}^{r}
\left[ 
{\left(
\hat{y}_k+L_k (\hat{\mathbf{p}})
\right)
}\rho 
{\left(
\hat{y}_k+L_k (\hat{\mathbf{p}})
\right)^{\dagger}}
\right.}
\\
%\shoveright
{
\left.
- \frac{1}{2}
{\left\{
{\left(
\hat{y}_k+L_k (\hat{\mathbf{p}})
\right)^{\dagger}}
{\left(
\hat{y}_k+L_k (\hat{\mathbf{p}})
\right)} , \rho
\right\}}
\right]}
\end{multline}
and
 \begin{multline}
 \label{p}
 \mathcal{L}_{P}[\rho ]=
 \int_{\mathbb{R}^3} \sum_{j=1}^{\infty}
 \left[
 U  (\mathbf{q})
 L_j(\mathbf{q},\hat{\mathbf{p}})\rho L^{\dagger}_j(\mathbf{q},\hat{\mathbf{p}})
 U ^{\dagger} (\mathbf{q}) 
 \right.
 \\ 
% \nonumber
 \shoveright{\left.  
 -
         \frac 12
         \left \{
         L^{\dagger}_j(\mathbf{q},\hat{\mathbf{p}})L_j(\mathbf{q},\hat{\mathbf{p}}),\rho  
         \right \}
 \right]
 d\mu (\mathbf{q})
 }
  \\
  %\nonumber
  \shoveleft{{}+
  \int_{\mathbb{R}^3} \sum_{j=1}^{\infty}
  \left[
  \omega_j (\mathbf{q})
  (U  (\mathbf{q})\rho L^{\dagger}_j(\mathbf{q},\hat{\mathbf{p}})U^{\dagger}  (\mathbf{q})-\rho L^{\dagger}_j(\mathbf{q},\hat{\mathbf{p}}))
  \right.}
  \\
 % \nonumber
  \shoveright{\left.
  {}
  +
  (U (\mathbf{q})L_j(\mathbf{q},\hat{\mathbf{p}})\rho U^{\dagger}  (\mathbf{q})-L_j(\mathbf{q},\hat{\mathbf{p}})\rho )\omega_j^{*} (\mathbf{q})
  \right]
  d\mu (\mathbf{q})
  }
 \\
 +
 \int_{\mathbb{R}^3} \sum_{j=1}^{\infty}
 \left[
 U  (\mathbf{q})\rho U^{\dagger}  (\mathbf{q})
 -\rho 
 -\frac{i}{\hbar}
 \frac{[\mathbf{q}\cdot\hat{\mathbf{x}},\rho]}{1+|\mathbf{q}|^2}
 \right]
 |\omega_j (\mathbf{q})|^2
 d\mu (\mathbf{q})
 \end{multline}
respectively, where
\begin{displaymath}
   \hat{y}_k
=\sum_{i=1}^{3}a_{ki}\hat{x}_i \qquad
  k=0,\ldots,r\> {\leq 3} \qquad a_{ki}\in \mathbb{R}
\end{displaymath}
are linear combinations of the three position operators $\hat{x}_i$, $L_k
(\hat{\mathbf{p}})$ and $L_j(\mathbf{q},\hat{\mathbf{p}})$ are generally complex functions of the momentum operators,
\begin{displaymath}
   U  (\mathbf{q}) =   e^{i\mathbf{q}\cdot\hat{\mathbf{x}}}
\end{displaymath}
are unitary operators corresponding to a translation in momentum or
boost, satisfying together with~\eqref{a} the Weyl form of the canonical commutation rules
\begin{equation}
   \label{weyl}
U (\mathbf{a})U  (\mathbf{q}) =e^{i\mathbf{q}\cdot{\mathbf{a}}}U  (\mathbf{q}) U (\mathbf{a}),
\end{equation}
$\omega_j (\mathbf{q})$ complex functions and $\mu
(\mathbf{q})$ a positive measure on $\mathbb{R}^3$ satisfying the L\'evy condition
\begin{displaymath}
  \int_{\mathbb{R}^3}\,  \frac{|\mathbf{q}|^2}{1+|\mathbf{q}|^2}
  \sum_{j=1}^{\infty}|\omega_j (\mathbf{q})|^2d\mu (\mathbf{q})<+\infty.
\end{displaymath}
As it appears this is a quite rich structure, allowing for the
description of a very broad class of physical phenomena, only having
in common invariance under translations. A first application has
already been mentioned above when considering the master equation for
quantum Brownian motion, describing the motion of a quantum test
particle in a gas, close to thermal equilibrium, which corresponds to
a particular realization of the Gaussian component~\eqref{g}. Further
examples related to more recent research work will be considered in
the next section.

\subsection{Translation-covariant mappings for the description of dissipation and decoherence}
\label{sec:transl-covar-mapp}

In the previous Section we have given the general expression of a
translation-covariant generator of a quantum-dynamical semigroup
according to Holevo's results. This operator expression can be
actually distinguished in two parts, as also happens in the classical
L\'evy-Khintchine formula, providing the general characterization of
the exponent of the characteristic function of a classical L\'evy
process (see e.g.\cite{Breuer2007} for a concise presentation from a
physicist's standpoint or\cite{Feller1971} for a more thorough
probabilistic treatment). Very roughly speaking in the presence of
translation covariance the dynamics can be essentially described in
terms of momentum exchanges between test particle and reservoir. The
Gaussian part corresponds to a situation in which the dynamics is
determined by a very large number of very small momentum transfers,
which in the case of finite variance leads to a Gaussian
process. E. g. in the case of quantum Brownian motion the test
particle is close to thermal equilibrium, so that typical values of
its momentum are much bigger than that of the gas particles due to its
bigger mass, and the momentum changes imparted in the single
collisions are therefore of relatively small amount. The Poisson part
on the contrary can account for a situation in which few interaction
events corresponding to significant momentum transfers drive the
dynamics, as happens for example in experiments on collisional
decoherence, where very few kicks already lead to a significant loss
of coherence. In addition to this the general expression can also
account for peculiar situations, typical of L\'evy processes, where
variance or mean of the momentum transfers do diverge, so that rare
but extremely strong events can give the predominant contributions to
the dynamics. We will now try to exemplify such situations referring
to recent research work, thus showing how paying attention to
covariance properties of dynamical mappings does not only lead to a
better and deeper understanding of known results as for the case of
damped harmonic oscillator, two-level system and quantum Brownian motion briefly
considered in Sect.\ref{sec:shift-covar-damp}, Sect.\ref{sec:rotat-covar-two} and
Sect.\ref{sec:transl-covar-quant} respectively, but also provides
important insights for the treatment of more complicated problems,
allowing for a unified description of apparently quite different situations.

\subsubsection{Dissipation and quantum linear Boltzmann equation}
\label{sec:dissipation-qlbe}

The well-known quantum Brownian motion master equation~\eqref{qbm}
provides as we have seen an example of realization of the Gaussian
component~\eqref{g} of the general structure of generators of
translation-covariant quantum-dynamical semigroups specified above. A
further example involving the Poisson component~\eqref{p} can be given
still considering the reduced dynamics of the centre of mass of a
test particle interacting through collisions with a gas,
however not focussing on the case of a very massive particle close to
thermal equilibrium, so that momentum transfers and therefore energy
transfers due to collision events between test particle and gas are
not necessarily small anymore. This kinetic stage of dynamical
description asks for a quantum version of the classical linear
Boltzmann equation, the equation being linear in the sense that the
gas is supposed to be and remain in equilibrium, so that only the
state of the test particle evolves in time. Such a master equation has
been recently obtained and its expression in the case in which the
scattering cross-section describing the collisions between test
particle and gas only depends on the transferred momentum
$\mathbf{q}$, which is always true in Born approximation, is given by\cite{Vacchini2000a,Vacchini2001a,Vacchini2001b,Vacchini2002a}
\begin{multline}
   \label{qlbe}
   \frac{d\rho}{dt}=
        {}-
        {i \over \hbar}[
        H_0
        ,
        \rho
        ]
+
        {2\pi \over\hbar}
        (2\pi\hbar)^3
        n
        \int d^3\!
        \mathbf{q}
        \,  
        {
        | \tilde{t} (q) |^2
        }
\\
        \Biggl[
        e^{\frac{i}{\hbar}\mathbf{q}\cdot\hat{\mathbf{x}}}
        \sqrt{
        S(\mathbf{q},E (\mathbf{q},\hat{\mathbf{p}}))
        }
        \rho
        \sqrt{
        S(\mathbf{q},E (\mathbf{q},\hat{\mathbf{p}}))
        }
        e^{-\frac{i}{\hbar}\mathbf{q}\cdot\hat{\mathbf{x}}}
        -
        \frac 12
        \left \{
        S(\mathbf{q},E (\mathbf{q},\hat{\mathbf{p}})),
        \rho
        \right \}
        \Biggr],
\end{multline}
where $n$ is the gas density, $\tilde{t} ({q})$ is the Fourier transform
of the interaction potential, $S$ a two-point correlation function
known as dynamic structure factor, depending on momentum transfer
$\mathbf{q}$ and energy transfer $E ({\mathbf{q},\mathbf{p}})$
\begin{displaymath}
   E (\mathbf{q},\mathbf{p})=\frac{(\mathbf{p}+\mathbf{q})^2}{2M}-\frac{\mathbf{p}^2}{2M}.
\end{displaymath}
The dynamic structure factor for a free gas of particles obeying
Maxwell-Boltzmann statistics has the explicit expression
\begin{displaymath}
         S_{\rm \scriptscriptstyle MB}(\mathbf{q},E)
        =
        \sqrt{\frac{\beta m}{2\pi}}        
        {
        1
        \over
        q
        }
       e^{-{
        \beta
        \over
             8m
        }
        {
        (2mE + q^2)^2
        \over
                  q^2
        }},
\end{displaymath}
while the general definition of dynamic structure factor reads
\begin{displaymath}
   S (\mathbf{q},E)=\frac{1}{2\pi\hbar}
\int dt \int d^3 \! \mathbf{x} \, 
        e^ {\frac{i}{\hbar}(E t -
        \mathbf{q}\cdot\mathbf{x})}
\frac{1}{N}\int d^3 \! \mathbf{y} \,
        \left \langle  
         N(\mathbf{y})  
         N(\mathbf{x}+\mathbf{y},t)
         \right \rangle,
\end{displaymath}
that is to say it is the Fourier transform with respect to momentum
and energy transfer of the two-point density-density correlation
function of the medium, $N(\mathbf{y})$ being the number density
operator of the gas. This
indirectly tells us that the dynamics of the test particle is indeed
driven by the density fluctuations in the fluid, due to its discrete
microscopic nature\cite{Petruccione2005a}.
In the structure of the master equation the unitary operators
$e^{\frac{i}{\hbar}\mathbf{q}\cdot\hat{\mathbf{x}}}$ and
$e^{-\frac{i}{\hbar}\mathbf{q}\cdot\hat{\mathbf{x}}}$ appearing to left
and right of the statistical operator do account for the momentum transfer
imparted to the test particle as a consequence of a certain collision;
the rate with which collisions characterized by a certain momentum
transfer happen however, do depend on the actual momentum of the test
particle described by the operator $\hat{\mathbf{p}}$ through the dependence
of the dynamic structure factor $S$ on $\mathbf{p}$. This mechanism
accounts for the approach to equilibrium, favouring collisions driving
the kinetic energy of the test particle towards the equipartition
value. The result can also be generalized to an arbitrary scattering
cross-section, not necessarily only depending on momentum transfer, by
introducing in the master equation instead of the scattering
cross-section an operator-valued scattering amplitude, averaged over
the gas particles momenta\cite{Hornberger2006b}. One can indeed check
that the quantum linear Boltzmann equation~\eqref{qlbe} and its
generalizations, apart from being in Lindblad form and manifestly
covariant, do admit the correct stationary state~\eqref{sqbm} and
drive the kinetic energy to its equipartition value.

\subsubsection{Decoherence and L\'evy processes}
\label{sec:decoh-levy-proc}

In the present Section we will show how translation-covariant
quantum-dynamical semigroups can provide a unified theoretical
description of quite different decoherence experiments. At odds with
the previous Section we are not interested in the dynamics of the
momentum observable, decoherence due to spatial localization usually
takes place on a much shorter time scale than relaxation phenomena. We
therefore neglect in the general expressions~\eqref{g} and~\eqref{p}
of a translation-covariant generator the dependence on the momentum
operator, which we take as a classical label, characterized e.g. by
the mean momentum of the incoming test particle. The
formulas~\eqref{g} and~\eqref{p} drastically simplify to
\begin{align}
\label{gpdecoh}
\mathcal{L}_{G}[\rho]&=-i 
 \sum_{i=1}^{3}
 {b}_i \left[  
         \hat{x}_i,\rho
         \right]  
 -\frac{1}{2}\sum_{i,j=1}^{3}{{D}_{ij}} \left[  
         \hat{x}_i,
         \left[  
         \hat{x}_j,\rho
         \right]  
         \right]  
 \\
 \mathcal{L}_{P}[\rho]&=
 \int d\mu (\mathbf{q}) 
  |\lambda(\mathbf{q})|^2
 \left[e^{\frac{i}{\hbar}\mathbf{q}\cdot \hat{\mathbf{x}}}\rho e^{-\frac{i}{\hbar}\mathbf{q}\cdot \hat{\mathbf{x}}}-\rho-\frac{i}{\hbar}
 \frac{[\mathbf{q}\cdot \hat{\mathbf{x}},\rho]}{1+|\mathbf{q}|^2}\right],
 \end{align}
with
\begin{displaymath}
   {b}_i\in\mathbb{R},\quad  {D}_{ij} \geq 0,\quad |\lambda(\mathbf{q})|^2=\sum_{j=1}^{\infty}|L^{\dagger}_j(\mathbf{q}) + \omega_j (\mathbf{q})|^2
\end{displaymath}
whose matrix elements in the position representation simply read
\begin{displaymath}
    \langle
    \mathbf{x}|\mathcal{L}_{G}[\rho]+\mathcal{L}_{P}[\rho]|\mathbf{y}\rangle 
 =-\Psi(\mathbf{x}-\mathbf{y})
 \langle \mathbf{x}|\rho|\mathbf{y}\rangle
\end{displaymath}
with
\begin{multline}
\label{cexp}
\Psi(\mathbf{x})=i
 \sum_{i=1}^{3}
 {b}_i {x}_i+\frac{1}{2}\sum_{i=1}^{3} {D}_{ij} {x}_i{x}_j
 \\
    -\int d\mu (\mathbf{q})  |\lambda(\mathbf{q})|^2
 \left[ e^{\frac{i}{\hbar}\mathbf{q}\cdot \mathbf{x}} -1
       -\frac{i}{\hbar}\frac{\mathbf{q}\cdot\mathbf{x}}{1+|\mathbf{q}|^2}\right],
 \end{multline}
which is exactly the general expression of the characteristic exponent
of a classical L\'evy process\cite{Feller1971,Breuer2007}. Neglecting the  free contribution the equation
for the time evolution of the statistical operator in the position
representation, which now becomes
\begin{displaymath}   
   \frac{d}{dt}\langle\mathbf{x}|\rho|\mathbf{y}\rangle=-\Psi(\mathbf{x}-\mathbf{y}) \langle
   \mathbf{x}|\rho|\mathbf{y}\rangle ,
\end{displaymath}
has solution
\begin{equation}
\label{decoh}
   \langle
   \mathbf{x}|\rho_t|\mathbf{y}\rangle =e^{ -t \Psi(\mathbf{x}-\mathbf{y})
   }\langle \mathbf{x}|\rho_0|\mathbf{y}\rangle
\equiv \Phi (t,\mathbf{x}-\mathbf{y})\langle \mathbf{x}|\rho_0|\mathbf{y}\rangle
\end{equation}
given by multiplying the matrix elements of the initial statistical
operator by the characteristic function
\begin{displaymath}
   \Phi (t,\mathbf{x})=e^{ -t \Psi(\mathbf{x})
   }
\end{displaymath}
of a classical L\'evy process evaluated at a point $(\mathbf{x}-\mathbf{y})$
given by the difference between the two spatial locations
characterizing bra and ket with which the matrix elements of $\rho_t$
are taken. In view of the general properties of the characteristic
function $\Phi$, which is the Fourier transform of a probability density, that is to say
\begin{align}
\label{cfp}
  & \Phi (t,0)=1
&&|\Phi (t,\mathbf{x}-\mathbf{y})|\leq 1 
\\
&\Phi
   (t,\mathbf{x}-\mathbf{y})
\xrightarrow{(\mathbf{x}-\mathbf{y}) \rightarrow \infty} 0
&&\Phi
   (t,\mathbf{x}-\mathbf{y})
\xrightarrow{t \rightarrow \infty} 0
\nonumber
\end{align}
the solution~\eqref{decoh} actually predicts on general grounds an exponential loss
of coherence in position, that is to say diagonalization in the
localization basis. In fact according to~\eqref{cfp} diagonal matrix
elements are left untouched by the dynamics, which together with the
fact that a characteristic function is actually a positive definite
function accounts for the correct probability and positivity preserving
time evolution. For growing time off-diagonal matrix elements are
fully suppressed, whatever the distance, while for fixed time
evolution the reduction of off-diagonal matrix elements depends on the
relative distance $(\mathbf{x}-\mathbf{y})$, leading to a vanishing
contribution for macroscopic distances (provided the corresponding
classical L\'evy process admits a proper probability density). An
application of this general theoretical treatment to actual physical
systems and in particular to relevant experimental situations relies
on a choice of functions and parameters appearing in~\eqref{cexp}
dictated by actual physical input. This has been accomplished
in\cite{Vacchini2005a}, where this general scheme has been connected
to actual experiments on decoherence, as well as theoretical
predictions of decoherence effects when the reservoir inducing
decoherence is a quantum chaotic system.

\subsubsection{Acknowledgements}
\label{sec:acknowledgements}

The author would like to thank Prof. Francesco Petruccione for the
invitation to hold these lectures, and the Centre for Quantum
Technology in Durban for financial support. He is very grateful to
Prof. L.  Lanz for careful reading of the manuscript and most useful
discussions on the subject of the paper. This research work was also
supported by MIUR under PRIN05.

%
%
% BibTe\rho users please use

 %   \bibliographystyle{klausarticle}
%    \bibliography{bassano}

%
% Non-BibTe\rho users please follow the syntax
% the syntax of "referenc.tex" for your own citations
% \begin{thebibliography}{99.}
% \end{thebibliography}
%%%%%%%%%%%%%%%%%%%%%%%%%%%%%%%%%%%%%%%%%%%%%%%%%%%%%%%%%%%%%%%%%%%%%%  }

%%%%%%%%%%%%%%%%%%%%%%%%%%%%%%%%%%%%%%%%%%%%%%%%%%%%%%%%%%%%%%%%%%%%%%

\printindex
\end{document}